\documentclass[%
reprint,
superscriptaddress,
amsmath,amssymb,
aps,
floatfix,
]{revtex4-1}

\usepackage{graphicx}
\usepackage{subcaption}
\usepackage{dcolumn}
\usepackage{bm}
\usepackage{braket}
\usepackage{amsmath}
\usepackage{mathtools}
\usepackage[colorlinks,allcolors=blue]{hyperref} 
\usepackage{hyperref}

\newcommand{\expect}[1]{\left\langle #1 \right\rangle}

\newcommand{\Schrodinger}{Schr\"{o}dinger }

\begin{document}
	
	\raggedbottom
	
	\preprint{APS/123-QED}
	
	\title{Optimal Quantum Transfer from Input Flying Qubit to Lossy Quantum Memory}
	
	\author{Eric Chatterjee}
	\affiliation{Sandia National Laboratories, Livermore, California 94550, USA}
	\author{Daniel Soh}
	\affiliation{Sandia National Laboratories, Livermore, California 94550, USA}
	\author{Matt Eichenfield}
	\affiliation{Sandia National Laboratories, Albuquerque, New Mexico 87123, USA}

	\date{\today}
	
	\begin{abstract} 
		In a quantum network, a key challenge is to minimize the direct reflection of flying qubits as they couple to stationary, resonator-based memory qubits, as the reflected amplitude represents state transfer infidelity that cannot be directly recovered. Optimizing the transfer fidelity can be accomplished by dynamically varying the resonator's coupling rate to the flying qubit field. Here, we analytically derive the optimal coupling rate profile in the presence of intrinsic loss of the quantum memory using an open quantum systems method that can account for intrinsic resonator losses. We show that, since the resonator field must be initially empty, an initial amplitude in the resonator must be generated in order to cancel reflections via destructive interference; moreover, we show that this initial amplitude can be made sufficiently small as to allow the net infidelity throughout the complete transfer process to be close to unity. We then derive the time-varying resonator coupling that maximizes the state transfer fidelity as a function of the initial population and intrinsic loss rate, providing a complete protocol for optimal quantum state transfer between the flying qubit and resonator qubit. We present analytical expressions and numerical examples of the fidelities for the complete protocol using exponential and Gaussian profiles. We show that a state transfer fidelity of around 99.9\% can be reached for practical intrinsic losses of resonators used as quantum memories.
	\end{abstract}
	
	\pacs{Valid PACS appear here}
	\maketitle
	
\section{Introduction} \label{sec: Introduction}

In a quantum network, the transfer of quantum information may occur frequently between stationary memory qubits and flying qubits that connect them. In the optical domain, a number of recent innovations have enabled the construction of resonator-based memories with tunable coupling rates  \cite{Tunableresonator1,Tunableresonator2,Tunableresonator3,Tunableresonator4}, and this allows for the possibility of dynamically tuning those coupling rates to maximize the fidelity of state transfer in a quantum network that uses those resonators as quantum memories. When a flying qubit containing quantum network information is incident on a resonator, its reflection from the resonator contributes to infidelity of state transfer together with intrinsic losses in the resonator. The reflection contains two independent contributions: a direct reflection from the resonator and leakage out of the resonator from the internal resonator field. The two contributions interfere and thus allow for the possibility that destructive interference can be used to minimize reflection and the resulting infidelity. The reflectivity of the resonator is intrinsically tied to its coupling rate, and thus building up amplitude in the internal resonator field while minimizing reflection by modifying the coupling rate is a non-trivial endeavor. Moreover, how one does this depends explicitly on the temporal profile of the flying qubit. Nevertheless, this is exactly what must be done in order to operate quantum networks with high fidelity and low error rates.  

\par 
Previous analysis has built on the idea of converting a propagating signal to resonator mode oscillations. Harlow considered the transfer from a propagating photon field to a stored phonon field via an optomechanical oscillator and used a semiclassical method to derive the optimal photon-phonon coupling coefficient over time in the oscillator \cite{HarlowThesis}. By contrast, we employ a fully-quantum method and specifically consider a two-mode system consisting of the propagating qubit and the resonator qubit. To this end, Wenner et al. \cite{Wenner} tuned the input signal to achieve near-unity fidelity. However, our goal is to maximize the fidelity for a generic input profile, which must be accomplished by adjusting the only tunable parameter, i.e. the coupling rate for the resonator. As a matter of fact, memory blocks with adjustable input-output coupling rate can be built in practice. For example, a two-sided resonator where one input-output interface is connected to a phase shifter and a high-reflective mirror has the required capability of high-speed output coupling rate adjustment \cite{Taylor}. Here, the adjustment of the input-output coupling is accomplished by changing the amount of phase shifting. Particularly, the round-trip phase shift $\pi$ corresponds to zero output coupling rate while the zero round-trip phase shift corresponds to maximum output coupling rate. Therefore, one can continuously adjust the output coupling rate between zero and maximum physically achievable value.

The most recent advancement was made by Nurdin et al. \cite{Nurdin}, who provided a full-quantum solution for an ideal, lossless resonator. Here, however, we aim to incorporate the nonidealities in the resonator that result in intrinsic loss, specifically deriving the transfer fidelity as a function of the loss rate. Furthermore, it is crucial to note that in order for the loss due to reflection to from the resonator to be zeroed out, there must be a non-zero population in the resonator field. This is due to the fact that the reflected signal can only be cancelled via destructive interference with the output from the resonator. Therefore, the overall transfer process requires a lossy initial step, which was treated numerically by Nurdin et al. In our analysis, however, we will analytically derive the parameters that optimize this lossy initial stage.
\par
Input-output theory for quantum systems, originally devised by Gardiner and Collett \cite{GardinerCollett}, provides a concise method for relating the input operators, output operators, and memory mode operators for an open quantum system. At a conceptual level, the net output for the system is generated from the interference between the directly scattered (reflected) part of the input signal and the coherent output from the system mode. If the coupling rate is tunable, then we can adjust it over time to evolve with the time-varying input field and resonator population, such that the directly scattered (reflected) input signal and the memory mode output are destructively interfering with each other at any given time during the transfer, resulting in zero net output through the coherent loss channel. The result of such zero output is apparently the perfect quantum transfer from the flying input qubit to the memory mode.
\par 
Although the coherent output serves as the main loss channel, it is also important to consider the intrinsic losses inherent in any system. Such intrinsic losses occur in any quantum memory regardless it is a ferminionic field (such as atomic energy states) or a bosonic field (such as a resonator field). Even for a well-manufactured optical cavity, for example, the high-reflectivity mirror can never quite reach 100 percent reflectivity; the state-of-the-art Bragg mirrors feature a minimum transmittance on the order of one ppm \cite{BraggMirrorsMinimumTransmittance}. Furthermore, it is tremendously important to recognize that the zero net output can only be achieved if there exists some initial population in the resonator such that the output from the resonator mode can destructively interfere with the reflected input signal. As a result, it will be unavoidable for a quantum memory with a vacuum initial condition to have the zero-output condition broken for a very short period of time, during which the necessary initial population is generated. Consequently, it is not completely possible to attain a perfect quantum transfer between an input flying qubit and the quantum memory mode. 
\par 
Here, we will analytically derive the optimal input coupling rate as a function of time for a given temporal profile of the input flying qubit. To do so, we model the input signal as originating from an imaginary resonator. We assign a matching time-varying output-coupling of this imaginary resonator to produce the originally given input flying qubit's temporal profile. It is worth noting that, for the case of zero intrinsic loss, our result reduces to that calculated by Nurdin et al. \cite{Nurdin}. We further verify the result for zero intrinsic loss using a semiclassical method. As an additional consideration, we treat the issue of generating the initial resonator population in an analytical manner, deriving the optimal amount of time over which the zero-output condition should be broken and the output coupling rate for the quantum memory resonator during that time period. We show that, even with this lossy first phase, we still attain a near perfect fidelity for an ideal resonator provided that the input rate is much lower than the maximum physically achievable output coupling rate. Lastly, we calculate the fidelity of the quantum transfer from the input signal to a nonideal resonator as a function of the resonator intrinsic loss for exponential and Gaussian input profiles, with the result showing that a fidelity in the range of 99.9\% can be attained for a Gaussian input to a practical resonator with an intrinsic loss rate 4 orders of magnitude below the maximum achievable output coupling rate.

\section{Full-Quantum Solution} \label{sec: Full-Quantum Solution}

The quantized field in the quantum memory can be either a two-level system (i.e., a qubit in an anharmonic system) or, more generally, an infinite-level boson field (i.e., a harmonic system). To treat the most general case, we assume that the quantum memory is comprised of a boson field in a cavity. We then consider a system consisting of a resonator and a propagating qubit encoded in the following superposition of vacuum and singly-excited states:
\begin{equation}
\ket{\psi} = c_g\ket{0} + c_e\ket{1}.
\end{equation}
For a given input probability current of the propagating qubit approaching the resonator, our goal is to optimize the quantum transfer from the input field to the resonator by minimizing the output field. The tunable parameter is the resonator output coupling rate $\kappa(t)$. Labeling the input field, output field, and resonator mode as $a_{in}$, $a_{out}$, and $a$, respectively, we apply the following input-output relationship \cite{GardinerCollett}:
\begin{equation} \label{eq: input-output}
a_{out}(t) = a_{in}(t) + \sqrt{\kappa(t)} a(t).
\end{equation}
In general, a bath input derives from a superposition of bath frequency modes, each of which interacts with the system via a frequency-dependent coupling coefficient. The picture is further complicated if the system mode is a harmonic ladder instead of a two-level mode. For the general case, the Hilbert space would be infinite-dimensional, consisting of the tensor product of the entire harmonic ladder for the resonator with the harmonic ladders corresponding to the infinite number of bath modes.
\par
Although a qubit can be realized in various manners, we specifically consider the case where the qubit consists of a superposition between a vacuum and a singly-excited state. This significantly simplifies the relevant bath Hilbert space, since we start with only 0 or 1 particle in the bath. Consequently, due to conservation of excitation number in a case of particle-exchange coupling (c.f., a beam-splitter interaction Hamiltonian), the resonator mode can only contain up to 1 particle, thus reducing even a harmonic ladder to a two-level mode. The Hilbert space becomes 3-dimensional, consisting of the composite vacuum state, a singly-excited-bath and vacuum-resonator state, and a singly-excited-resonator and vacuum-bath state. Although the bath particle belongs in a superposition of frequency modes, we envision that the bath mode is analogous to a two-level imaginary resonator since its occupation number is no greater than 1. 
\par
\begin{figure}[!tb]
	\centering
	\includegraphics[width=0.8\linewidth]{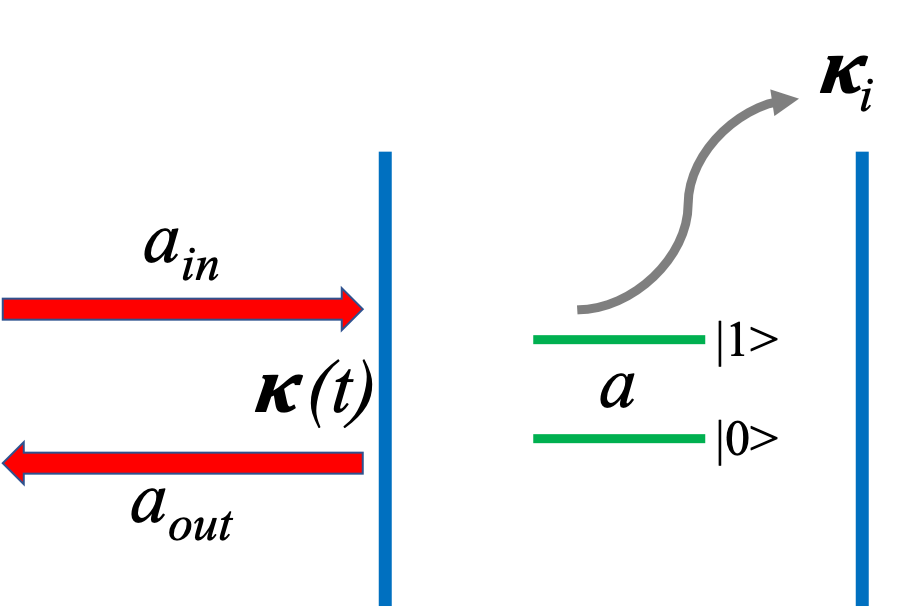}
	\caption{Depiction of quantum state transfer from the flying qubit to the resonator mode $a$. The operators $a_{in}$ and $a_{out}$ denote the resonator input and output, respectively; $\kappa_i$ represents the resonator's intrinsic loss rate; the time-adjustible coupling between the resonator and the flying qubit is given by $\kappa(t)$.}
	\label{fig:maindiagram}
\end{figure}

Figure~\ref{fig:maindiagram} depicts the quantum transfer from the propagating qubit to the resonator mode.
Labeling the annihilation operator of this first resonator mode as $a_1$, the time-varying output-coupling rate for that resonator as $\kappa_1(t)$, and the vacuum input field for the first resonator as $a_{in,1}$, the input field $a_{in}(t)$ for the second resonator is re-written as follows:
\begin{equation} \label{eq: ain(t)}
a_{in}(t) = a_{in,1} + \sqrt{\kappa_1(t)} a_1.
\end{equation}
Note that this expression will enable us to relate the abstract quantity $\kappa_1(t)$ to the given temporal profile of the input signal, which will be shown later in this section.
\par
Next, we construct the composite wavefunction in the Hilbert space composed of a tensor product of the two-level resonator subspaces, truncated to exclude the doubly-occupied state. The Hilbert space is thus spanned by the vacuum state $\ket{00}$ and the singly-excited states $(\ket{10},\ket{01})$, where $a_1$ and $a$ represent the annihilation operators for the first and second slots, respectively. We wish to transfer a qubit consisting of a superposition of $\ket{00}$ and $\ket{10}$ to an identical superposition of $\ket{00}$ and $\ket{01}$:
\begin{equation}
\ket{\Psi} = c_g \ket{00} + c_e \ket{10} \rightarrow c_g \ket{00} + c_e \ket{01}.
\end{equation}
Defining $\beta_1(t)$ and $\beta(t)$ as time-varying coefficients whose amplitude-squared values represent the populations of resonators 1 and 2 respectively, we find that the singly-excited part of the wavefunction evolves as follows:
\begin{equation} \label{eq: Psi(t)}
\ket{\Psi(t)} = c_g \ket{00} + c_e \Big(\beta_1(t) \ket{10} + \beta(t) \ket{01}\Big).
\end{equation}
Our goal is to zero out the output from the second resonator, which is accomplished by ensuring that $a_{out}(t)(\ket{\Psi(t)} \otimes \ket{\mathrm{vac}}) = 0$ where $\ket{\mathrm{vac}}$ is the vacuum bath input to the imaginary cavity. Synthesizing Eqs.~\eqref{eq: input-output},~\eqref{eq: ain(t)} and~\eqref{eq: Psi(t)}, and dividing both sides by $c_e$, we find that the output coupling rate $\kappa(t)$ for the second resonator must be designed such that the following condition holds:
\begin{equation} \label{eq: zero loss}
\sqrt{\kappa_1(t)} \beta_1(t) + \sqrt{\kappa(t)} \beta(t) = 0.
\end{equation}
It is also important to consider how we can relate the abstract quantity $\kappa_1(t)$, corresponding to the output coupling rate of the imaginary first resonator, to the given input rate (normalized with respect to the initial singly-excited population $|c_e|^2$) for the second resonator, which we label $r_{in}(t)$. Intuitively, we can deduce that the output coupling rate of the imaginary resonator should be directly proportional to the transfer rate of quanta from the imaginary to the quantum memory resonator and inversely proportional to the population of the imaginary resonator. To demonstrate this quantitatively, we relate $r_{in}(t)$ to $\kappa_1(t)$ and $\beta_1(t)$ by using the expression for $a_{in}(t)$ from Eq.~\eqref{eq: ain(t)} and taking the expectation value of both sides for the state $\ket{\Psi(t)}$:
\begin{equation} \label{eq: rin(t)}
	\kappa_1(t) = \frac{\expect{a_{in}^{\dag}(t) a_{in}(t)}}{\expect{a_1^{\dag} a_1}} = \frac{|c_e|^2 r_{in}(t)}{|c_e|^2 |\beta_1(t)|^2} = \frac{r_{in}(t)}{|\beta_1(t)|^2}. 
\end{equation}
We are now ready to introduce the Hamiltonian for the system and use the \Schrodinger equation to solve for the coefficients $\beta_1(t)$ and $\beta(t)$. Intuitively, we know that the interaction Hamiltonian between the populated bath mode and the resonator mode should take the form of a beam splitter. Since the rate of the interaction is proportional to $\kappa(t)$ as well as to the input rate from the bath (which in turn is proportional to $\kappa_1(t)$), we intuitively know that the interaction coefficient should be proportional to $\sqrt{\kappa_1(t)\kappa(t)}$. This is quantitatively confirmed by employing SLH formalism \cite{SLHFormalism}. In the rotating frame for the resonator modes, the SLH triple for the first resonator takes the following form:
\begin{equation}
S_1 = I \textrm{, } L_1 = \sqrt{\kappa_1(t)} a_1 \textrm{, } H_1 = 0.
\end{equation}
Analogously, the SLH triple for the second resonator is the following in the rotating frame:
\begin{equation}
S_2 = I \textrm{, } L_2 = \sqrt{\kappa(t)} a \textrm{, } H_2 = 0.
\end{equation}
Since the first and second resonators interact unidirectionally (i.e. the output of the first resonator serves as the input for the second), we use the series connection rule to derive the SLH triple for the composite system, labeling the first and second resonators as $C_1$ and $C_2$, respectively, and the composite system as $C_T$:
\begin{widetext}
\begin{align}
	(S_T,L_T,H_T) &= C_T = C_2 \triangleleft C_1 = \bigg(S_2S_1, L_2 + S_2L_1, H_1 + H_2 - \frac{i\hbar}{2}\Big(L_2^{\dag}S_2L_1 - L_1^{\dag}S_2^{\dag}L_2\Big)\bigg) \nonumber  \\
	&= \bigg(I, \sqrt{\kappa(t)}a + \sqrt{\kappa_1(t)}a_1, \frac{i\hbar}{2} \sqrt{\kappa_1(t) \kappa(t)} \Big(a_1^{\dag}a - a_1a^{\dag}\Big)\bigg).
\end{align}
\end{widetext}
Note that the composite Lindbladian $L_T$ is equivalent to the output operator $a_{out}$ for the second resonator, which we wish to zero out by properly designing $\kappa(t)$. As a result, in this lossless system, the Hermitian Hamiltonian $H_T$ fully determines the time-evolution, and the time evolution is now unitary. 
\par
So far, we have assumed a perfect quantum memory having only one input-output coupling channel. However, in reality, additional undesired loss channels may exist. These may include the spontaneous emission of atomic systems or the intrinsic loss of a resonator. Labeling the aggregated intrinsic loss rate as $\kappa_i$, the following Lindbladian represents this decoherence process:
\begin{equation}
L_i = \sqrt{\kappa_i} a.
\end{equation}
Since the system can contain no more than one excitation, the time-evolution of the Hilbert subspace consisting of singly-excited states (represented by the coefficients $c_e\beta_1(t)$ and $c_e\beta(t)$) can be modeled by adding an anti-Hermitian (and thus lossy) term to the Hamiltonian (see Appendix~\ref{sec: Master Equation to Non-Hermitian Hamiltonian}):
\begin{align}
	H_T^\mathrm{eff} \ket{\Psi(t)} &= \bigg(H_T - \frac{i\hbar}{2} L_i^{\dag} L_i\bigg) \ket{\Psi(t)} \nonumber \\
	&= \bigg(H_T - i\hbar \frac{\kappa_i}{2} a^{\dag} a\bigg) \ket{\Psi(t)}.
\end{align}
Since this effective Hamiltonian is non-Hermitian, the time evolution of the wavefunction is non-unitary, i.e. $\beta_1^2(t) + \beta^2(t) \ne 1$. We must therefore solve separately for $\beta_1(t)$ and $\beta(t)$, and we start by solving the \Schrodinger equation row-by-row for this effective Hamiltonian:
\begin{align}
i\hbar c_e \dot{\beta_1}(t) &= \braket{10|H_T^\mathrm{eff}|\Psi(t)} \nonumber \\
&= \frac{i\hbar}{2} \sqrt{\kappa_1(t) \kappa(t)} c_e \beta(t), \\
i\hbar c_e \dot{\beta}(t) &= \braket{01|H_T^\mathrm{eff}|\Psi(t)} \nonumber  \\
&= -\frac{i\hbar}{2} \sqrt{\kappa_1(t) \kappa(t)} c_e \beta_1(t) - i\hbar \frac{\kappa_i}{2} c_e \beta(t).
\end{align}
Dividing both expressions by $i\hbar c_e$, we find a system of differential equations corresponding to the time-evolution of the state coefficients:
\begin{align} 
\dot{\beta_1}(t) &= \frac{\sqrt{\kappa_1(t) \kappa(t)}}{2} \beta(t), \label{eq: dot beta1(t)} \\
\dot{\beta}(t) &= -\frac{\sqrt{\kappa_1(t) \kappa(t)}}{2} \beta_1(t) - \frac{\kappa_i}{2} \beta(t). \label{eq: dot beta(t)} 
\end{align}
Note that the coefficients in these differential equations are fully real. Therefore, given real initial values $\beta_1(t_i)$ and $\beta(t_i)$, the coefficients will remain real-valued throughout the time evolution. Furthermore, if we successfully design the output coupling temporal profile for the real second resonator such that the reflected signal from that resonator is zeroed out, we intuitively deduce that the dynamics of the imaginary first resonator should be invariant in the dynamics of the second, since the connection between the two resonators is fully mediated through the output from the first resonator. We demonstrate this quantitatively by substituting the desired input-output relationship from Eq.~\eqref{eq: zero loss} into Eq.~\eqref{eq: dot beta1(t)}:
\begin{align}
	\dot{\beta_1}(t) &= \frac{\sqrt{\kappa_1(t)}}{2} \Big(\sqrt{\kappa(t)} \beta(t)\Big) = \frac{\sqrt{\kappa_1(t)}}{2} \Big(-\sqrt{\kappa_1(t)} \beta_1(t)\Big) \nonumber  \\
	&= -\frac{\kappa_1(t)}{2} \beta_1(t).
\end{align}
Substituting Eq.~\eqref{eq: rin(t)} into Eq.~\eqref{eq: dot beta1(t)}, we re-write the differential equation for $\dot{\beta_1}(t)$ as a function of $r_{in}(t)$ instead of $\kappa_1(t)$, thus abstracting out the imaginary resonator:
\begin{align} \label{eq: dot beta1(t) alternate}
	\dot{\beta_1}(t) = -\frac{1}{2} \frac{r_{in}(t)}{\beta_1^2(t)} \beta_1(t) = -\frac{r_{in}(t)}{2\beta_1(t)}.
\end{align}
We can use this expression to derive the time-evolution of the imaginary resonator population in terms of the given normalized transfer rate $r_{in}(t)$ from the first to the second resonator, provided that the output coupling profile for the second cavity is designed to zero out the reflected signal. To do so, we multiply both sides by $dt2\beta_1$ and integrate to find $\beta_1^2(t)$:
\begin{align} \label{eq: beta_1(t) squared}
	\int_{\beta_1(t_i)}^{\beta_1(t)} d\beta_1 2\beta_1 &= -\int_{t_i}^t dt r_{in}(t), \nonumber  \\
	\beta_1^2(t) &= \beta_1^2(t_i) - \int_{t_i}^t dt r_{in}(t).
\end{align}
Next, we turn our attention to Eq.~\eqref{eq: dot beta(t)}. We aim to express the first term on the right hand side in terms of $\beta(t)$ rather than $\beta_1(t)$. To begin, we abstract out the properties of the first resonator by replacing $\kappa_1(t)$ and $\beta_1(t)$ with a function of the input rate $r_{in}(t)$, for which we take the square root of Eq.~\eqref{eq: rin(t)}:
\begin{equation} \label{eq: sqrt rin(t)}
\sqrt{\kappa_1(t)} \beta_1(t) = \pm \sqrt{r_{in}(t)}.
\end{equation}
Note the $\pm$ sign in front of the right-hand-side term. Since $\sqrt{r_{in}(t)}$ and $\sqrt{\kappa_1(t)}$ are always non-negative, the sign on the right-hand-side will be identical to the sign of $\beta_1(t)$.
\par
Next, we express $\sqrt{\kappa(t)}$ in terms of $r_{in}(t)$ and $\beta(t)$. From the input-output relationship in Eq.~\eqref{eq: input-output} and the square root of Eq.~\eqref{eq: rin(t)} shown in Eq.~\eqref{eq: sqrt rin(t)}, we find the following:
\begin{equation} \label{eq: sqrt kappa(t)}
	\sqrt{\kappa(t)} = -\frac{\sqrt{\kappa_1(t)} \beta_1(t)}{\beta(t)} = \mp \frac{\sqrt{r_{in}(t)}}{\beta(t)}.
\end{equation}
Substituting Eqs.~\eqref{eq: sqrt rin(t)} and~\eqref{eq: sqrt kappa(t)} into Eq.~\eqref{eq: dot beta(t)}, we find that $\dot{\beta(t)}$ is now entirely a function of the dynamics of the second resonator (namely $\beta(t)$, the input rate $r_{in}(t)$, and the intrinsic loss rate $\kappa_i$), as desired:
\begin{align}
	\dot{\beta}(t) &= -\frac{1}{2} \Big(\pm \sqrt{r_{in}(t)}\Big) \Bigg(\mp \frac{\sqrt{r_{in}(t)}}{\beta(t)}\Bigg) - \frac{\kappa_i}{2} \beta(t) \nonumber  \\
	&= \frac{r_{in}(t)}{2\beta(t)} - \frac{\kappa_i}{2} \beta(t).
\end{align}
We are now in a position to solve for $\beta(t)$. Multiplying both sides by $2\beta(t)$, the expression becomes a first-order ordinary differential equation in $\beta^2(t)$:
\begin{equation}
	\frac{d}{dt} \beta^2(t) = 2\beta(t)\dot{\beta}(t) = r_{in}(t) - \kappa_i \beta^2(t).
\end{equation}
Intuitively, this expression corresponds to the fact that the rate of change of the second (real) resonator population equals the input rate minus the loss rate through the intrinsic channel. We integrate this using the standard formula for first-order ODEs:
\begin{equation}
\beta^2(t) = e^{-\kappa_i(t - t_i)} \bigg(\int_{t_i}^t dt e^{\kappa_i(t - t_i)} r_{in}(t) + G\bigg),
\end{equation}
where $G$ is a constant. By using the example of $t = t_i$, it becomes apparent that $G = \beta^2(t_i)$, representing the initial population of the second resonator:
\begin{equation} \label{eq: beta^2(t)}
\beta^2(t) = e^{-\kappa_i(t - t_i)} \bigg(\beta^2(t_i) + \int_{t_i}^t dt e^{\kappa_i(t - t_i)} r_{in}(t)\bigg).
\end{equation}
Lastly, we solve for the optimal output coupling rate profile $\kappa(t)$ by synthesizing Eqs.~\eqref{eq: zero loss},~\eqref{eq: rin(t)}, and~\eqref{eq: beta^2(t)}:
\begin{align} \label{eq: kappa(t)}
	\kappa(t) &= \frac{\kappa_1(t) \beta_1^2(t)}{\beta^2(t)} \nonumber  \\
	&= \frac{r_{in}(t)}{e^{-\kappa_i(t - t_i)} \Big(\beta^2(t_i) + \int_{t_i}^t dt e^{\kappa_i(t - t_i)} r_{in}(t)\Big)}.
\end{align}
Note that $\kappa(t)$ is independent of the dynamics of the imaginary resonator (i.e. population and output coupling rate), instead varying only with the given input rate profile for the quantum memory resonator, as desired.

\section{Generating an Initial Population} \label{sec: Generating an Initial Population}

Based on the result from Eq.~\eqref{eq: kappa(t)}, it is evident that, in order for the coherent output from the quantum memory resonator to equal zero, a non-zero initial population is required--otherwise, the initial optimal output coupling rate would be infinite. At a qualitative level, we know that the zero net output condition is satisfied if the directly scattered (reflected) input signal from a resonator interferes destructively with the output from the resonator mode. If the resonator mode is empty, then there does not exist any mode output that can destructively interfere with the reflected signal. As such, it is necessary to set the coupling to some value $\kappa(t_i)$ that breaks the zero-output requirement for a very short period of time $\delta t$ such that a small initial population can be generated. In order to determine the constraints on $\kappa(t_i)$ and $\delta t$, we re-solve the \Schrodinger equation for $t = t_i$ to $t = t_i + \delta t$, this time incorporating the Lindbladian corresponding to the coherent output as an anti-Hermitian term in the effective Hamiltonian (see Appendix~\ref{sec: Master Equation to Non-Hermitian Hamiltonian}):
\begin{widetext}
\begin{align}
H_T^\mathrm{eff} &= \frac{i\hbar}{2} \sqrt{\kappa_1(t) \kappa(t)} \Big(a_1^{\dag}a - a_1a^{\dag}\Big) - \frac{i\hbar}{2} \Big(\sqrt{\kappa(t)}a^{\dag} + \sqrt{\kappa_1(t)}a_1^{\dag}\Big) \Big(\sqrt{\kappa(t)}a + \sqrt{\kappa_1(t)}a_1\Big) - i\hbar \frac{\kappa_i}{2} a^{\dag}a.
\end{align}
\end{widetext}
It is worth considering why the coherent output serves as a loss channel from the Hilbert space. If we had two quantum memory resonators connected in series, with the direction of propagation perpendicular to the resonator mirrors, then we would expect the output signal from the second resonator to return to the first as an input. However, in the present model, the first resonator actually represents a bath mode. Therefore, the output signal does not interact directly with the input signal. This results in a unidirectional propagation from the first to the second resonator regardless of the size of the second resonator output, as quantitatively demonstrated by expanding $H_T^\mathrm{eff}$:
\begin{align}
H_T^\mathrm{eff} &= -\frac{i\hbar}{2} \Big(2\sqrt{\kappa_1(t) \kappa(t)}a_1a^{\dag} + \kappa(t)a^{\dag}a + \kappa_1(t)a_1^{\dag}a_1\Big) \nonumber \\
&\quad - i\hbar \frac{\kappa_i}{2} a^{\dag}a.
\end{align}
With the Hamiltonian in this form, we see that the terms in $a_1^{\dag}a$ have cancelled out, while the terms in $a_1a^{\dag}$ remain, corresponding to the unidirectionality of the transfer. As in Section~\ref{sec: Full-Quantum Solution}, we solve the \Schrodinger equation row-by-row:
\begin{align}
i\hbar c_e \dot{\beta_1}(t) &= \braket{10|H_T^\mathrm{eff}|\Psi(t)} = -\frac{i\hbar}{2} \kappa_1(t) c_e \beta_1(t), \\
i\hbar c_e \dot{\beta}(t) &= \braket{01|H_T^\mathrm{eff}|\Psi(t)} \nonumber \\
&= -i\hbar \sqrt{\kappa_1(t) \kappa(t)} c_e \beta_1(t) - \frac{i\hbar}{2} \Big(\kappa(t) + \kappa_i\Big) c_e \beta(t).
\end{align}
Dividing both sides by $i\hbar c_e$:
\begin{align} 
\dot{\beta_1}(t) &= -\frac{\kappa_1(t)}{2} \beta_1(t), \label{eq: dot beta1(t) full} \\
\dot{\beta}(t) &= -\sqrt{\kappa_1(t) \kappa(t)} \beta_1(t) - \frac{\kappa(t) + \kappa_i}{2} \beta(t). \label{eq: dot beta(t) full}
\end{align}
Next, we solve for the reflectional loss rate for a generic output coupling rate $\kappa(t)$. To start, we use the time evolution of the coefficients $\beta_1$ and $\beta$ to determine the time evolution of the populations $\beta_1^2$ and $\beta^2$:
\begin{align} \label{eq: dot beta1sq(t) full}
\begin{split}
\dot{\beta_1^2}(t) &= 2\beta_1(t)\dot{\beta_1}(t) \\
&= -\kappa_1(t) \beta_1^2(t) \\
&= -r_{in}(t),
\end{split}
\end{align}
\begin{align} \label{eq: dot betasq(t) full}
\begin{split}
\dot{\beta^2}(t) &= 2\beta(t)\dot{\beta}(t) \\ &= -2\sqrt{\kappa_1(t) \kappa(t)} \beta_1(t) \beta(t) - \Big(\kappa(t) + \kappa_i\Big) \beta^2(t) \\
&= -2 \sqrt{r_{in}(t) \kappa(t)} \beta(t) - \Big(\kappa(t) + \kappa_i\Big) \beta^2(t)
\end{split}
\end{align}
Note that the final step in Eq.~\eqref{eq: dot betasq(t) full} is based on the relationship $\sqrt{r_{in}(t)} = \sqrt{\kappa_1(t)} \beta_1(t)$. The lack of ambiguity in the sign is made possible by the fact that $\beta_1(0)$ always equals 1, and the time evolution governed by Eq.~\eqref{eq: dot beta1sq(t) full} ensures that $\beta_1(t)$ remains positive for all times $t$. Based on energy conservation, we know that the input signal entering the resonator at rate $r_{in}$ equals the sum of the rate of filling of the resonator ($\dot{\beta^2}$), the loss rate  through the input-output channel due to reflection (which we label $r_{out}$), and the intrinsic loss rate ($\kappa_i \beta^2$). The input-output loss rate $r_{out}$ is thus expressed as follows:
\begin{align} \label{eq: r_out}
\begin{split}
r_{out} &= r_{in} - \dot{\beta^2} - \kappa_i \beta^2 \\
&= \beta^2(t) \kappa(t) + 2\sqrt{r_{in}(t)} \beta(t) \sqrt{\kappa(t)} + r_{in}(t) \\
&= \Big(\beta(t)\sqrt{\kappa(t)} + \sqrt{r_{in}(t)}\Big)^2.
\end{split}
\end{align}
It is evident that this is a quadratic equation in $\sqrt{\kappa(t)}$. Furthermore, since $\beta_1(t)$ is always positive, the time-evolution of $\beta(t)$ as governed by Eq.~\eqref{eq: dot beta(t) full} ensures that $\beta(t)$ is negative for all $t > 0$. As such, the input-output loss rate is zeroed out when $\kappa(t) = r_{in}(t)/\beta^2(t)$, confirming the result of the previous section. For very low resonator population $\beta^2(t)$, however, this can only be accomplished if $\kappa(t)$ is greater than the maximal physically achievable output coefficient $\kappa_{max}$. Therefore, for this initial stage in which we generate enough population in the resonator such that the zero-output condition can be satisfied thereafter, we set $\kappa(t) = \kappa_{max}$ in order to minimize the input-output loss rate $r_{out}$:
\begin{equation}
r_{out} = \Big(\beta(t)\sqrt{\kappa_{max}} + \sqrt{r_{in}(t)}\Big)^2.
\end{equation}
Minimizing the loss rate through the input-output channel not only minimizes the rate of fidelity degradation during the initial stage, but it also minimizes the time elapsed until sufficient resonator population is built up such that the zero-loss stage can begin. Intuitively, this is due to the fact that at any given time for any given input rate, a minimal input-output loss rate implies a maximal filling rate for the resonator. This is quantitatively demonstrated by the the first line of Eq.~\eqref{eq: r_out}, which shows that for a fixed input rate $-\dot{\beta_1^2}$, a minimal value for $r_{out}$ implies a maximal value for $\dot{\beta^2} + \kappa_i \beta^2$ and thus a maximal value for $\beta^2(t)$ at all times $t$. We therefore conclude that setting $\kappa(t) = \kappa_{max}$ for the initial stage maximizes the overall fidelity in this stage by minimizing both the input-output loss rate and the time over which this loss is incurred.
\par 
We now proceed to deriving the resonator population $\beta^2$ as a function of time $t$ for this initial stage, given a resonator output coupling rate $\kappa(t) = \kappa_{max}$. It is convenient to start by solving for $\beta(t)$ from Eq.~\eqref{eq: dot beta(t) full}, which we re-write in terms of the input rate $r_{in}(t)$:
\begin{equation}
\dot{\beta}(t) = -\sqrt{\kappa_{max}} \sqrt{r_{in}(t)} - \frac{\kappa_{max} + \kappa_i}{2} \beta(t)
\end{equation}
In general, the intrinsic loss coefficient $\kappa_i$ for the resonator will be negligible compared to the maximum achievable output coupling rate $\kappa_{max}$, enabling us to use the approximation $\kappa_{max} + \kappa_i \approx \kappa_{max}$. Nonetheless, for the sake of accuracy, we maintain the original form and solve for $\beta(t)$ through the standard treatment for first-order ordinary differential equations:
\begin{align} \label{eq: beta(t) first stage}
\begin{split}
\beta(t) &= e^{-\frac{\kappa_{max} + \kappa_i}{2}t} \int_0^t dt e^{\frac{\kappa_{max} + \kappa_i}{2}t} \Big(-\sqrt{\kappa_{max}} \sqrt{r_{in}(t)}\Big).
\end{split}
\end{align}
The resonator population $\beta^2(t)$ is calculated trivially by squaring this expression:
\begin{align} \label{eq: beta^2(t) first stage}
\begin{split}
\beta^2(t) &= \kappa_{max} e^{-(\kappa_{max} + \kappa_i) t} \bigg(\int_0^t dt e^{\frac{\kappa_{max} + \kappa_i}{2}t} \sqrt{r_{in}(t)}\bigg)^2.
\end{split}
\end{align}
From this, we can determine the time required for this stage as well as the total input-output loss. As previously discussed in this section, the input-output loss rate is zero when $\kappa(t) = r_{in}(t)/\beta^2(t)$. Therefore, the threshold resonator population is reached when $\beta^2(t_c) = r_{in}(t_c)/\kappa_{max}$, which occurs at the time we define as $t = t_c$. We substitute this into Eq.~\eqref{eq: beta^2(t) first stage}:
\begin{align} \label{eq: t_c relationship}
\begin{split}
&\frac{r_{in}(t_c)}{\kappa_{max}} \\
&= \kappa_{max} e^{-(\kappa_{max} + \kappa_i) t_c} \bigg(\int_0^{t_c} dt e^{\frac{\kappa_{max} + \kappa_i}{2}t} \sqrt{r_{in}(t)}\bigg)^2, \\
&\frac{\sqrt{r_{in}(t_c)}}{\kappa_{max}} e^{\frac{\kappa_{max} + \kappa_i}{2}t_c} = \int_0^{t_c} dt e^{\frac{\kappa_{max} + \kappa_i}{2}t} \sqrt{r_{in}(t)}.
\end{split}
\end{align}
Given an input temporal profile $r_{in}$, we will substitute into this expression and solve for $t_c$. Then, for the second (zero input-output loss) stage, the time evolution of the resonator population is calculated from Eq.~\eqref{eq: beta^2(t)} using an initial value of $\beta^2(t_c)$, which in turn is determined by substituting $t = t_c$ in Eq.~\eqref{eq: beta^2(t) first stage}:
\begin{widetext}
\begin{align} \label{eq: beta^2(t) second stage}
\begin{split}
\beta^2(t) &= e^{-\kappa_i(t - t_c)} \bigg(\beta^2(t_c) + \int_{t_c}^t dt e^{\kappa_i(t - t_c)} r_{in}(t)\bigg) \\
&= e^{-\kappa_i t} \Bigg(\kappa_{max} e^{-\kappa_{max} t_c} \bigg(\int_0^{t_c} dt e^{\frac{\kappa_{max} + \kappa_i}{2}t} \sqrt{r_{in}(t)}\bigg)^2 + \int_{t_c}^t dt e^{\kappa_i t} r_{in}(t)\Bigg) \\
&= e^{-\kappa_i t} \Bigg(\frac{r_{in}(t_c)}{\kappa_{max}} e^{\kappa_i t_c} + \int_{t_c}^t dt e^{\kappa_i t} r_{in}(t)\Bigg),
\end{split}
\end{align}
\end{widetext}
where the last line is derived by re-writing the first integral by substituting the relationship in Eq.~\eqref{eq: t_c relationship}. The optimal output coupling profile $\kappa(t)$ for $t > t_c$ is calculated by dividing the input rate by the population, per Eq.~\eqref{eq: kappa(t)}:
\begin{align} \label{eq: kappa(t) second stage}
\begin{split}
\kappa(t) &= r_{in}(t) e^{\kappa_i t} \Bigg(\frac{r_{in}(t_c)}{\kappa_{max}} e^{\kappa_i t_c} + \int_{t_c}^t dt e^{\kappa_i t} r_{in}(t)\Bigg)^{-1}.
\end{split}
\end{align}

\section{Sample Profiles and Fidelity} \label{sec: Sample Profiles and Fidelity}

It is useful to calculate the transfer fidelity and required output coupling profiles $\kappa(t)$ for sample normalized input rate profiles $r_{in}(t)$. Since the actual input rate $|c_e|^2 r_{in}(t)$ results in a final occupation probability of $|c_e|^2$ for the resonator (in the absence of loss), the normalized input signal must carry a total of 1 excitation. Therefore, the following condition must be satisfied:
\begin{equation}
\int_{t_i}^{t_f} r_{in}(t) dt = 1.
\end{equation}
As discussed in the previous sections, the transfer process occurs in two stages. The first takes place from $t = t_i$ to $t = t_i + \delta t$, during which enough initial population in the resonator is generated such that the output from the resonator can thereafter cancel the reflected input signal through destructive interference. In this stage, the non-zero loss through the input-output channel serves as the main loss channel, while the intrinsic loss is very low in comparison due to the small value of $\delta t$. The second phase takes place from $t = t_i + \delta t$ to $t = t_f$, during which the output coupling rate is dynamically tuned so that the input-output loss is zero, resulting in intrinsic loss serving as the sole loss channel. Here, we will numerically calculate the fidelity after stage 2 as a function of the ratio between the resonator intrinsic loss rate $\kappa_i$ and the initial input rate $r_{in}(t_i)$.
\par
We analyze two shapes for the input intensity profile: exponentially decreasing and Gaussian. For each of the input shapes, we minimize the loss in the first stage by setting the output coupling rate to $\kappa(t_i) = \kappa_{max}$ (see the previous section).
\par

\subsection{Exponentially decaying input}

For a given initial input rate $r$, the input shape $r_{in}(t)$ in the case of an exponentially decaying input must take the following form in order for the total signal to equal unity (single-qubit input condition):
\begin{equation} \label{eq: exponential input}
r_{in}(t) = r e^{-rt}.
\end{equation}
We start by solving for the time $t_c$ at which we switch from the lossy initial stage to the zero-input-output-loss stage. Substituting Eq.~\eqref{eq: exponential input} into Eq.~\eqref{eq: t_c relationship} yields the following expression:
\begin{align}
\begin{split}
\frac{\sqrt{r}}{\kappa_{max}} &e^{\frac{(\kappa_{max}+\kappa_i-r)}{2}t_c} = \sqrt{r} \int_0^{t_c} dt e^{\frac{(\kappa_{max}+\kappa_i-r)}{2}t}, \\
&t_c = \frac{2}{\kappa_{max} + \kappa_i - r} \ln{\bigg(\frac{2\kappa_{max}}{\kappa_{max} - \kappa_i + r}\bigg)}.
\end{split}
\end{align}
It is convenient to convert time to dimensionless units by expressing it in multiples of $1/\kappa_{max}$. As such, we define dimensionless quantities $\tau$ and $\tau_c$, corresponding to $\kappa_{max} t$ and $\kappa_{max} t_c$, respectively. We can then demonstrate that $\tau_c$ is a function specifically of the intrinsic loss rate $\kappa_i$ and maximum input rate $r$ \textit{as fractions of} $\kappa_{max}$:
\begin{align}
\begin{split}
\tau_c &= \frac{2 \kappa_{max}}{\kappa_{max} + \kappa_i - r} \ln{\bigg(\frac{2\kappa_{max}}{\kappa_{max} - \kappa_i + r}\bigg)} \\
&= \frac{2}{1 + \frac{\kappa_i}{\kappa_{max}} - \frac{r}{\kappa_{max}}} \ln{\Bigg(\frac{2}{1 - \frac{\kappa_i}{\kappa_{max}} + \frac{r}{\kappa_{max}}}\Bigg)}.
\end{split}
\end{align}
Next, we focus on calculating the resonator population and optimal output coupling profile for $t > t_c$. To start, we evaluate the integrals in Eqs.~\eqref{eq: beta^2(t) second stage} and~\eqref{eq: kappa(t) second stage} given an exponential input profile. For the first integral expression, this is accomplished by substituting Eq.~\eqref{eq: t_c relationship}:
\begin{align}
\begin{split}
&\bigg(\int_0^{t_c} dt e^{\frac{\kappa_{max} + \kappa_i}{2}t} \sqrt{r_{in}(t)}\bigg)^2 \\
&= \frac{r_{in}(t_c)}{\kappa_{max}^2} e^{(\kappa_{max} + \kappa_i)t_c} \\
&= \frac{r}{\kappa_{max}^2} e^{(\kappa_{max} + \kappa_i - r)t_c} \\
&= \frac{r}{\kappa_{max}^2} \bigg(\frac{2\kappa_{max}}{\kappa_{max} - \kappa_i + r}\bigg)^2 \\
&= \frac{4r}{(\kappa_{max} - \kappa_i + r)^2}.
\end{split}
\end{align}
The second integral (proportional to the gain in resonator population from time $t_c$ to an arbitrary $t$, where $t > t_c$) is solved through direct integration:
\begin{align}
\begin{split}
\int_{t_c}^t dt e^{\kappa_i t} r_{in}(t) &= r \int_{t_c}^t dt e^{(\kappa_i - r)t} \\
&= \frac{r}{r - \kappa_i} \Big(e^{(\kappa_i - r)t_c} - e^{(\kappa_i - r)t}\Big) \\
&= \frac{r}{r - \kappa_i} \Bigg(\bigg(\frac{2\kappa_{max}}{\kappa_{max} - \kappa_i + r}\bigg)^{\frac{2(\kappa_i - r)}{\kappa_{max} + \kappa_i - r}} \\
&\quad - e^{(\kappa_i - r)t}\Bigg).
\end{split}
\end{align}
We are now in a position to solve for the resonator population $\beta^2(t)$ and the optimal output coupling profile $\kappa(t)$ for $t > t_c$. The population is calculated by substituting the above expressions into Eq.~\eqref{eq: beta^2(t) second stage}:
\begin{widetext} \label{eq: beta^2(t) exponential input}
\begin{align}
\begin{split}
\beta^2(t) &= e^{-\kappa_i t} \Bigg(e^{-\kappa_{max} t_c} \frac{4r\kappa_{max}}{(\kappa_{max} - \kappa_i + r)^2} + \frac{r}{r - \kappa_i} \Big(e^{(\kappa_i - r)t_c} - e^{(\kappa_i - r)t}\Big)\Bigg) \\
&= e^{-\kappa_i t} \Bigg(\frac{4r\kappa_{max}}{(\kappa_{max} - \kappa_i + r)^2} \bigg(\frac{2\kappa_{max}}{\kappa_{max} - \kappa_i + r}\bigg)^{-\frac{2\kappa_{max}}{\kappa_{max} + \kappa_i - r}} + \frac{r}{r - \kappa_i} \bigg(\frac{2\kappa_{max}}{\kappa_{max} - \kappa_i + r}\bigg)^{\frac{2(\kappa_i - r)}{\kappa_{max} + \kappa_i - r}} - \frac{r}{r - \kappa_i} e^{(\kappa_i - r)t}\Bigg),
\end{split}
\end{align}
\end{widetext}
and the optimal output coupling profile is calculated from Eq.~\eqref{eq: kappa(t) second stage}:
\begin{widetext}
\begin{align}
\begin{split}
\kappa(t) &= re^{-rt} e^{\kappa_i t} \Bigg(e^{-\kappa_{max} t_c} \frac{4r\kappa_{max}}{(\kappa_{max} - \kappa_i + r)^2} + \frac{r}{r - \kappa_i} \Big(e^{(\kappa_i - r)t_c} - e^{(\kappa_i - r)t}\Big)\Bigg)^{-1} \\
&= r e^{(\kappa_i - r)t} \Bigg(\frac{4r\kappa_{max}}{(\kappa_{max} - \kappa_i + r)^2} \bigg(\frac{2\kappa_{max}}{\kappa_{max} - \kappa_i + r}\bigg)^{-\frac{2\kappa_{max}}{\kappa_{max} + \kappa_i - r}} + \frac{r}{r - \kappa_i} \bigg(\frac{2\kappa_{max}}{\kappa_{max} - \kappa_i + r}\bigg)^{\frac{2(\kappa_i - r)}{\kappa_{max} + \kappa_i - r}} \\
&\quad - \frac{r}{r - \kappa_i} e^{(\kappa_i - r)t}\Bigg)^{-1}.
\end{split}
\end{align}
\end{widetext}
Returning to the conversion of time $t$ to the dimensionless quantity $\tau$, we can show that $\beta^2(\tau)$ and $\kappa(\tau)$ vary with $\kappa_i$ and $r$ only as fractions of $\kappa_{max}$, as was the case with $\tau_c$. We start by re-writing the expression for $\beta^2(\tau)$, where $\tau > \tau_c$:
\begin{align} \label{eq: beta^2(tau) exponential input}
\begin{split}
\beta^2(\tau) &= A_1 e^{-\frac{\kappa_i}{\kappa_{max}} \tau} - A_2 e^{-\frac{r}{\kappa_{max}} \tau},
\end{split}
\end{align}
where $A_1$ and $A_2$ are constants representing the following:
\begin{widetext}
\begin{align}
A_1 &= \frac{4r\kappa_{max}}{(\kappa_{max} - \kappa_i + r)^2} \bigg(\frac{2\kappa_{max}}{\kappa_{max} - \kappa_i + r}\bigg)^{-\frac{2\kappa_{max}}{\kappa_{max} + \kappa_i - r}} + \frac{r}{r - \kappa_i} \bigg(\frac{2\kappa_{max}}{\kappa_{max} - \kappa_i + r}\bigg)^{\frac{2(\kappa_i - r)}{\kappa_{max} + \kappa_i - r}} \\
&= \frac{4}{\Big(\sqrt{\frac{\kappa_{max}}{r}} - \sqrt{\frac{\kappa_i}{\kappa_{max}} \frac{\kappa_{max}}{r} \frac{\kappa_i}{\kappa_{max}}} + \sqrt{\frac{r}{\kappa_{max}}}\Big)^2} \Bigg(\frac{2}{1 - \frac{\kappa_i}{\kappa_{max}} + \frac{r}{\kappa_{max}}}\Bigg)^{-\frac{2}{1 + \frac{\kappa_i}{\kappa_{max}} - \frac{r}{\kappa_{max}}}} \\
&\quad + \frac{1}{1 - \frac{\kappa_i}{\kappa_{max}} \frac{\kappa_{max}}{r}} \Bigg(\frac{2}{1 - \frac{\kappa_i}{\kappa_{max}} + \frac{r}{\kappa_{max}}}\Bigg)^{\frac{2\Big(\frac{\kappa_i}{\kappa_{max}} - \frac{r}{\kappa_{max}}\Big)}{1 + \frac{\kappa_i}{\kappa_{max}} - \frac{r}{\kappa_{max}}}}, \\
A_2 &= \frac{r}{r - \kappa_i} \\
&= \frac{1}{1 - \frac{\kappa_i}{\kappa_{max}} \frac{\kappa_{max}}{r}}.
\end{align}
\end{widetext}
The output coupling profile $\kappa(\tau)$ for $\tau > \tau_c$ can be expressed in terms of the same dimensionless coefficients:
\begin{align} \label{eq: kappa(tau) exponential input}
\begin{split}
&\kappa(\tau) \\
&= \kappa_{max} \frac{r}{\kappa_{max}} e^{-\frac{r}{\kappa_{max}} \tau} \Big(A_1 e^{-\frac{\kappa_i}{\kappa_{max}} \tau} - A_2 e^{-\frac{r}{\kappa_{max}} \tau}\Big)^{-1} \\
&= \kappa_{max} \frac{r}{\kappa_{max}} \Bigg(A_1 e^{\Big(\frac{r}{\kappa_{max}} - \frac{\kappa_i}{\kappa_{max}}\Big) \tau} - A_2\Bigg)^{-1}.
\end{split}
\end{align}
For $\tau < \tau_c$, during which $\kappa(\tau) = \kappa_{max}$, we determine the time evolution of the resonator population by substituting the exponential input profile into Eq.~\eqref{eq: beta^2(t) first stage}:
\begin{widetext}
\begin{align}
\begin{split}
\beta^2(t) &= \kappa_{max} r e^{-(\kappa_{max} + \kappa_i) t} \bigg(\int_0^t dt e^{\frac{\kappa_{max} + \kappa_i -r}{2}t}\bigg)^2 \\
&= \kappa_{max} r \bigg(\frac{2}{\kappa_{max} + \kappa_i - r}\bigg)^2 e^{-(\kappa_{max} + \kappa_i) t} \bigg(e^{\frac{\kappa_{max} + \kappa_i - r}{2}t} - 1\bigg)^2 \\
&= \frac{4\kappa_{max}r}{(\kappa_{max} + \kappa_i - r)^2} \bigg(e^{-\frac{r}{2} t} - e^{-\frac{\kappa_{max} + \kappa_i}{2} t}\bigg)^2.
\end{split}
\end{align}
\end{widetext}
Conversion of time $t$ to the dimensionless quantity $\tau$ yields the following expression for $\beta^2(\tau)$ for $\tau < \tau_c$:
\begin{widetext}
\begin{equation}
\beta^2(\tau) = \frac{4}{\Big(\sqrt{\frac{\kappa_{max}}{r}} - \sqrt{\frac{\kappa_i}{\kappa_{max}} \frac{\kappa_{max}}{r} \frac{\kappa_i}{\kappa_{max}}} + \sqrt{\frac{r}{\kappa_{max}}}\Big)^2} \Bigg(e^{-\frac{1}{2}\frac{r}{\kappa_{max}} \tau} - e^{-\frac{1}{2} \Big(1 + \frac{\kappa_i}{\kappa_{max}}\Big) \tau}\Bigg)^2.
\end{equation}
\end{widetext}
Finally, we seek to determine the fidelity of the quantum state transfer from the input signal to the resonator. For an exponential input, the input rate drops over time. On the other hand, the intrinsic loss rate always rises in proportion to the resonator population, thus increasing over time. Given non-zero intrinsic loss, we intuitively deduce that the resonator population is maximized when the input rate drops below the intrinsic loss rate. The time $t_{max}$ at which this crossover occurs is determined quantitatively by taking the derivative of $\beta^2(t)$ from Eq.~\eqref{eq: beta^2(tau) exponential input} and setting it to zero, yielding the following result:
\begin{equation}
t_{max} = \frac{1}{r - \kappa_i} \ln{\bigg(\frac{r A_2}{\kappa_i A_1}\bigg)}.
\end{equation}
We convert this to the dimensionless variable $\tau_{max} = \kappa_{max} t_{max}$:
\begin{align}
\begin{split}
\tau_{max} &= \frac{\kappa_{max}}{r - \kappa_i} \ln{\bigg(\frac{r A_2}{\kappa_i A_1}\bigg)} \\
&= \frac{1}{\frac{r}{\kappa_{max}} - \frac{\kappa_i}{\kappa_{max}}} \ln{\bigg(\frac{r}{\kappa_{max}} \frac{\kappa_{max}}{\kappa_i} \frac{A_2}{A_1}\bigg)}.
\end{split}
\end{align}
The fidelity $F = \beta^2(\tau_{max})$ is thus calculated as follows:
\begin{align}
\begin{split}
&F \\
&= \beta^2(\tau_{max}) \\
&= A_1 \bigg(\frac{r A_2}{\kappa_i A_1}\bigg)^{-\frac{\kappa_i}{r - \kappa_i}} - A_2 \bigg(\frac{r A_2}{\kappa_i A_1}\bigg)^{-\frac{r}{r - \kappa_i}} \\
&= \bigg(\frac{r A_2}{\kappa_i A_1}\bigg)^{-\frac{r}{r - \kappa_i}} \Bigg(A_1 \bigg(\frac{r A_2}{\kappa_i A_1}\bigg)  - A_2\Bigg) \\
&= \bigg(\frac{\kappa_i}{\kappa_{max}} \frac{\kappa_{max}}{r} \frac{A_1}{A_2}\bigg)^{\frac{1}{1 - \frac{\kappa_i}{\kappa_{max}} \frac{\kappa_{max}}{r}}} A_2 \bigg(\frac{r}{\kappa_{max}} \frac{\kappa_{max}}{\kappa_i} - 1\bigg).
\end{split}
\end{align}
Note that for a resonator with zero intrinsic loss ($\kappa_i = 0$), the fidelity reduces to $F = A_1$, attained as $t \rightarrow \infty$, as expected since the resonator always experiences a net gain in population for any input rate in the absence of intrinsic loss. 
\par
\begin{figure}[!tb]
	\centering
	\includegraphics[width=\linewidth]{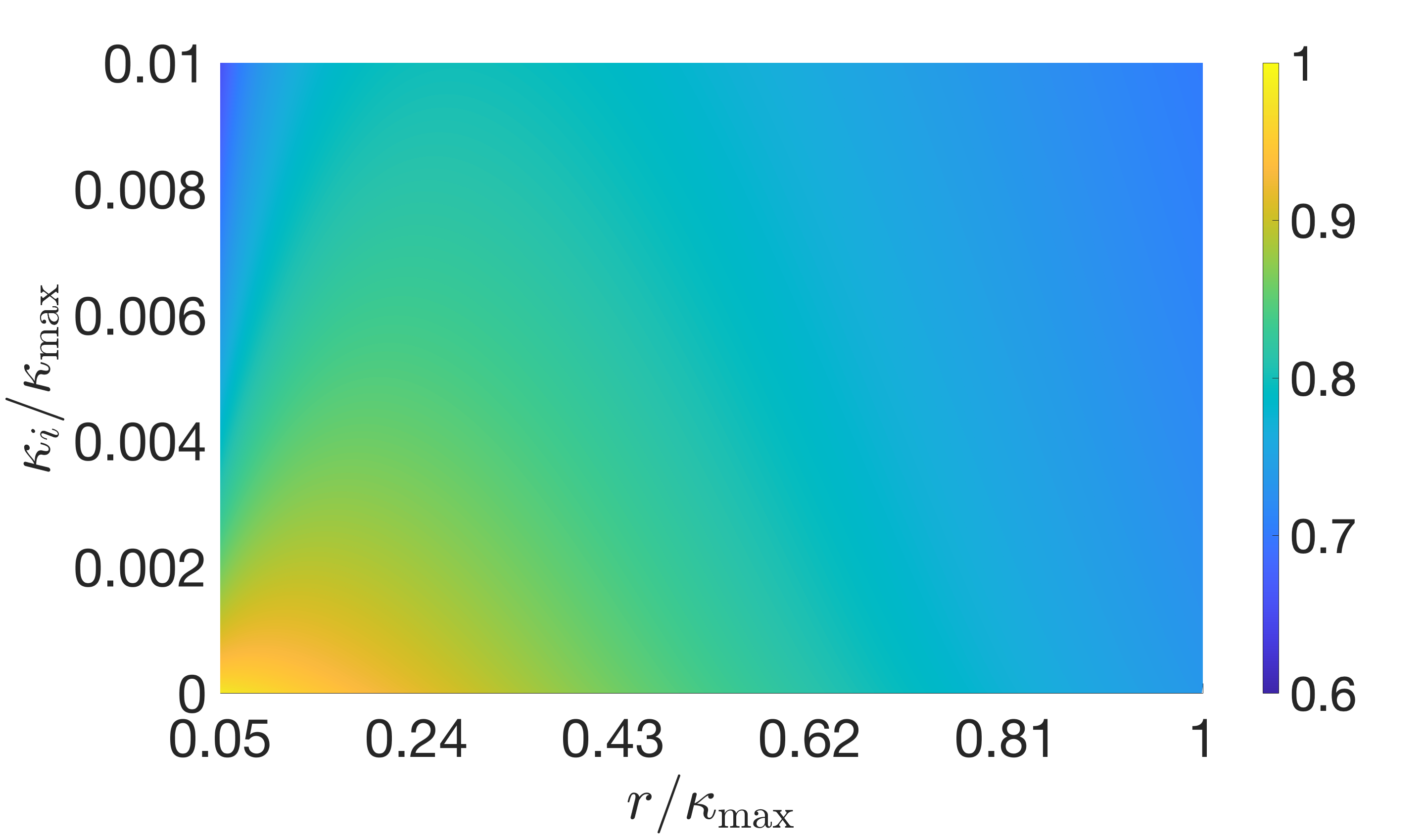}
	\caption{Fidelity $F$ as a function of intrinsic loss rate $\kappa_i$ and initial (maximum) input rate $r$, both normalized with respect to the maximal achievable output coupling rate $\kappa_\mathrm{max}$.}
	\label{fig:fidelityexponential}
\end{figure}

 
Figure~\ref{fig:fidelityexponential} depicts the fidelity as a function of the intrinsic loss rate $\kappa_i$ and the initial input rate $r$ in the range $\kappa_i \le \kappa_{max}/100$ and $\kappa_{max}/20 \le r \le \kappa_{max}$.
Analyzing the variation along each parameter, we note that the fidelity increases with decreasing intrinsic loss rate, as expected. However, the relationship with respect to the initial input rate is more intricate. Specifically, for a given intrinsic loss, the fidelity is lower at both the upper and lower bounds for $r$, while being maximized in the middle of the range. This can be attributed to the fact that for high input rate (relative to the maximum output coupling rate), the loss rate through the input-output channel during the initial stage is elevated, while for low input rate, the longer timescale for the transfer leads to greater loss through the intrinsic channel. We can explain the former dynamic intuitively as follows: A higher input rate leads to increased reflection from the resonator, thus necessitating a higher output coupling rate for the resonator in order for this reflected signal to be cancelled through destructive interference with the resonator output. As such, increasing the value of the input rate relative to the maximal achievable output coupling rate will lead to increased loss through the input-output channel during the first stage. As for the intrinsic loss, the intuition that longer input timescale leads to greater intrinsic loss also explains the fact that at higher intrinsic loss values, the optimal initial input rate becomes greater.
\par
\begin{figure*}[!tb]
	\centering
	\begin{subfigure}{\columnwidth}
		\centering
		\includegraphics[width=\linewidth]{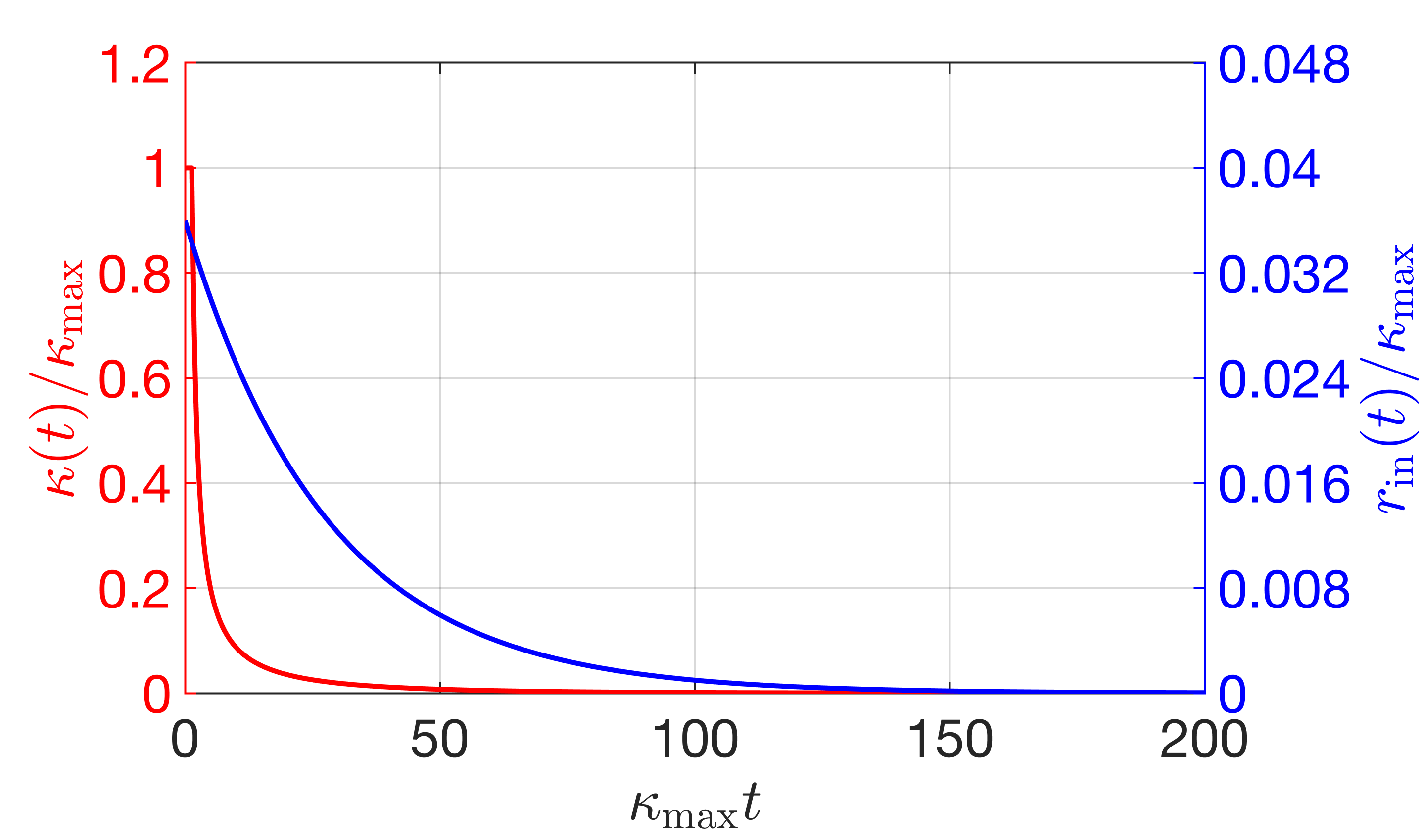}
		\caption{Input rate (blue) and optimal output coupling rate (red) vs. time.}
	\end{subfigure}
	\begin{subfigure}{\columnwidth}
		\centering
		\includegraphics[width=\linewidth]{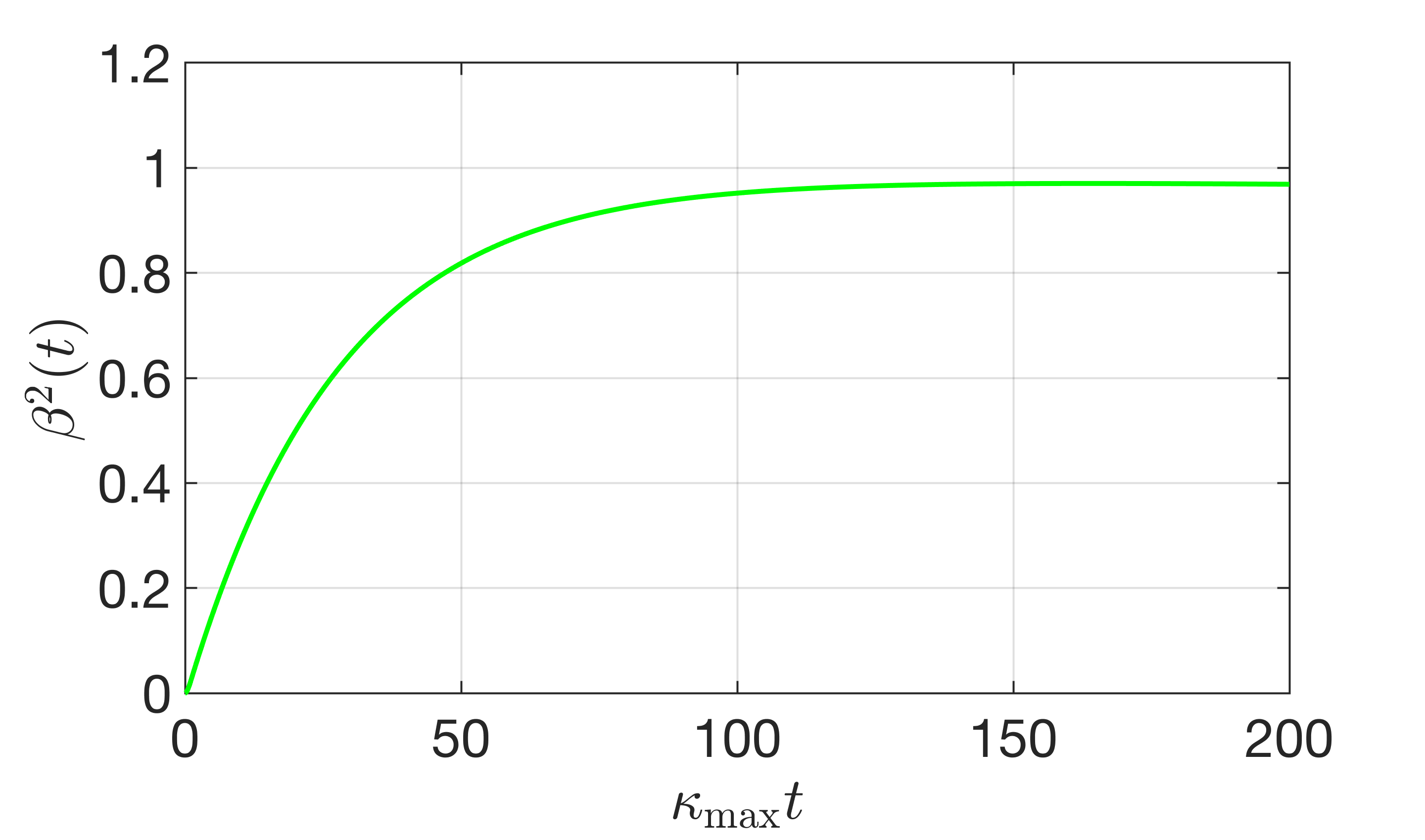}
		\caption{Resonator population vs. time.}
	\end{subfigure}
\caption{(Exponential input case) Optimal output coupling profile $\kappa(t)$ (a), and evolution of the resonator population $\beta^2(t)$ (b), given an intrinsic loss rate $\kappa_i = \kappa_\mathrm{max}/10000$ and initial (maximum) input rate $r = 0.036 \kappa_\mathrm{max}$.}
\label{fig:exponentialinput}
\end{figure*}

 
Figure~\ref{fig:exponentialinput} depicts the input profile, output coupling profile, and resonator population time-evolution for the selected parameters $\kappa_i = \kappa_{max}/10000$ and $r = 0.036 \kappa_{max}$. 
As desired, the optimal output coupling rate steadily decreases in the second stage. It is also apparent from the resonator population plot that the fidelity is 97\%, which is attained at time $t_{max} = 164.3/\kappa_{max}$.

\subsection{Gaussian input}

Next, we consider a Gaussian input profile. In general, for a standard deviation $\sigma$, the input temporal profile $r_{in}(t)$ is expressed in the following normalized manner:
\begin{equation} \label{eq: Gaussian input}
r_{in}(t) = \frac{1}{\sigma\sqrt{2\pi}} e^{-\frac{(t-t_0)^2}{2\sigma^2}},
\end{equation}
where $t_0$ denotes the time at which the input rate reaches its peak. In principle, $t_0$ equals infinity due to the fact that the width of the Gaussian is unlimited. It is thus necessary to set a start time for the absorption of the input. In order to ensure that a set percentage of the input signal is included in the range $t>0$ regardless of the width of the pulse, we make $t_0$ proportional to the standard deviation $\sigma$:
\begin{equation}
t_0 = n\sigma.
\end{equation}
It is desirable to satisfy the condition $n \gtrapprox 4$, since this ensures that at least 99.99\% of the input signal is covered.
\par 
As in the exponential case, we start by solving for the threshold time $t_c$ between the initial lossy stage and the zero input-output-loss stage. Substituting Eq.~\eqref{eq: Gaussian input} into Eq.~\eqref{eq: t_c relationship}, we obtain the following integral expression:
\begin{align}
\begin{split}
&e^{\frac{1}{2}\Big(1 + \frac{\kappa_i}{\kappa_{max}}\Big) \kappa_{max}t_c - \frac{\kappa_{max}^2 (t_c - t_0)^2}{4(\kappa_{max}\sigma)^2}} \\
&= \kappa_{max} \int_0^{t_c} dt e^{\frac{1}{2}\Big(1 + \frac{\kappa_i}{\kappa_{max}}\Big) \kappa_{max}t - \frac{\kappa_{max}^2 (t - t_0)^2}{4(\kappa_{max}\sigma)^2}}.
\end{split}
\end{align}
It is convenient to convert time to dimensionless units by defining a new variable $\tau = \kappa_{max} t$ (and consequently $\tau_0 = \kappa_{max} t_0$ and $\tau_c = \kappa_{max} t_c$). Then, all of the arguments in the above integral expression also become dimensionless:
\begin{align} \label{eq: t_c relationship Gaussian input}
\begin{split}
e^{a \tau_c - b(\tau_c - \tau_0)^2} = \int_0^{\tau_c} d\tau e^{a \tau - b(\tau - \tau_0)^2},
\end{split}
\end{align}
where $a$ and $b$ are functions of the intrinsic loss $\kappa_i$ and the input pulse standard deviation $\sigma$, respectively:
\begin{align}
a &= \frac{1}{2} \bigg(1 + \frac{\kappa_i}{\kappa_{max}}\bigg), \\
b &= \frac{1}{4(\kappa_{max} \sigma)^2}.
\end{align}
As the above relationships demonstrate, the threshold time $t_c$ is specifically a function of the intrinsic loss rate as a fraction of the maximum output coupling rate $\kappa_{max}$, as well as the peak input rate (proportional to $1/\sigma$) as a fraction of $\kappa_{max}$. Intuitively, we deduce that a higher intrinsic loss rate relative to $\kappa_{max}$ will extend the threshold time $t_c$ by delaying the filling of the resonator to the threshold population required for the second stage to begin. 
\par 
We now solve the integral in Eq.~\eqref{eq: t_c relationship Gaussian input}, which yields the following expression:
\begin{widetext}
\begin{equation} \label{eq: t_c Gaussian}
e^{a \tau_c - b(\tau_c - \tau_0)^2} = \frac{1}{2\sqrt{b}} e^{\frac{a^2}{4b} + a\tau_0} \sqrt{\pi} \Bigg(\mathrm{Erf}\bigg(\frac{a + 2b\tau_0}{2\sqrt{b}}\bigg) + \mathrm{Erf}\bigg(\frac{-a + 2b(\tau_c - \tau_0)}{2\sqrt{b}}\bigg)\Bigg).
\end{equation}
\end{widetext}
Since this expression cannot be analytically solved for the threshold time $\tau_c$, we instead design a numerical simulation in MATLAB, in which we loop over intrinsic loss values in the range $\kappa_i < \kappa_{max}$ and peak input rate values in the range $10\kappa_i < 1/\Big(\sigma\sqrt{2\pi}\Big) < \kappa_{max}$. Note that the latter condition corresponds to the following range for the dimensionless standard deviation $\kappa_{max}\sigma$:
\begin{equation}
\frac{1}{\sqrt{2\pi}} < \kappa_{max}\sigma < \frac{1}{10\sqrt{2\pi}} \bigg(\frac{\kappa_{max}}{\kappa_i}\bigg).
\end{equation}
Upon solving for $\tau_c$ for each value of intrinsic loss rate and input pulse standard deviation, we now focus on determining the time evolution of the resonator population $\beta^2(\tau)$ and the optimal output coupling rate temporal profile $\kappa(\tau)$ for the given pair of parameters. For $\tau > \tau_c$, $\beta^2$ is calculated by substituting into Eq.~\eqref{eq: beta^2(t) second stage}:
\begin{widetext}
\begin{align} \label{eq: beta^2(t) Gaussian input}
\begin{split}
\beta^2(\tau) &= e^{-\frac{\kappa_i}{\kappa_{max}} \tau} \Bigg(\frac{1}{\kappa_{max}} \frac{1}{\sigma\sqrt{2\pi}} e^{-\frac{(\tau_c - \tau_0)^2}{2(\kappa_{max}\sigma)^2}} e^{\frac{\kappa_i}{\kappa_{max}} \tau_c} + \frac{1}{\kappa_{max}} \frac{1}{\sigma\sqrt{2\pi}} \int_{\tau_c}^{\tau} d\tau e^{\frac{\kappa_i}{\kappa_{max}} \tau - \frac{(\tau - \tau_0)^2}{2(\kappa_{max}\sigma)^2}}\Bigg) \\
&= \frac{2\sqrt{b}}{\sqrt{2\pi}} e^{-(2a-1)\tau} \bigg(e^{(2a-1)\tau_c - 2b(\tau_c - \tau_0)^2} + \int_{\tau_c}^{\tau} d\tau e^{(2a-1)\tau - 2b(\tau - \tau_0)^2}\bigg),
\end{split}
\end{align}
\end{widetext}
where the integral evaluates to the following:
\begin{widetext}
\begin{equation}
\int_{\tau_c}^{\tau} d\tau e^{(2a-1)\tau - 2b(\tau - \tau_0)^2} = \frac{\sqrt{\pi}}{2\sqrt{2b}} e^{\frac{(2a-1)(2a+8b\tau_0-1)}{8b}} \Bigg(\mathrm{Erf}\bigg(\frac{1-2a+4b(\tau-\tau_0)}{2\sqrt{2b}}\bigg) - \mathrm{Erf}\bigg(\frac{1-2a+4b(\tau_c-\tau_0)}{2\sqrt{2b}}\bigg)\Bigg)
\end{equation}
\end{widetext}
Consequently, the optimal output coupling profile $\kappa(t)$ for $t > t_c$ is calculated from Eq.~\eqref{eq: kappa(t) second stage}:
\begin{widetext}
\begin{align}
\begin{split}
\kappa(t) &= \Bigg(\kappa_{max} \frac{2\sqrt{b}}{\sqrt{2\pi}} e^{-2b(\tau - \tau_0)^2}\Bigg) \frac{\sqrt{2\pi}}{2\sqrt{b}} e^{(2a-1)\tau} \bigg(e^{(2a-1)\tau_c - 2b(\tau_c - \tau_0)^2} + \int_{\tau_c}^{\tau} d\tau e^{(2a-1)\tau - 2b(\tau - \tau_0)^2}\bigg)^{-1} \\
&= \kappa_{max} e^{(2a-1)\tau - 2b(\tau - \tau_0)^2} \bigg(e^{(2a-1)\tau_c - 2b(\tau_c - \tau_0)^2} + \int_{\tau_c}^{\tau} d\tau e^{(2a-1)\tau - 2b(\tau - \tau_0)^2}\bigg)^{-1}.
\end{split}
\end{align}
\end{widetext}
For the first stage ($\tau < \tau_c$), over which $\kappa(\tau) = \kappa_{max}$, we solve for the resonator population time evolution by substituting the Gaussian input profile into Eq.~\eqref{eq: beta^2(t) first stage} and replacing $t$ with the dimensionless quantity $\tau$:
\\
\begin{align}
\begin{split}
&\beta^2(\tau) \\
&= \frac{e^{\frac{1}{2} \Big(1 + \frac{\kappa_i}{\kappa_{max}}\Big)\tau}}{\kappa_{max}} \Bigg(\int_0^{\tau} d\tau e^{\frac{1}{2} \Big(1 + \frac{\kappa_i}{\kappa_{max}}\Big)\tau} \sqrt{r_{in}(\tau)}\Bigg)^2 \\
&= \frac{e^{\frac{1}{2} \Big(1 + \frac{\kappa_i}{\kappa_{max}}\Big)\tau}}{\kappa_{max} \sigma\sqrt{2\pi}} \Bigg(\int_0^{\tau} d\tau e^{\frac{1}{2} \Big(1 + \frac{\kappa_i}{\kappa_{max}}\Big)\tau - \frac{(\tau - \tau_0)^2}{4(\kappa_{max}\sigma)^2}}\Bigg)^2 \\
&= \frac{2\sqrt{b}}{\sqrt{2\pi}} e^{-2a\tau} \bigg(\int_0^{\tau} d\tau e^{a \tau - b(\tau - \tau_0)^2}\bigg)^2,
\end{split}
\end{align}
where the integral expression is evaluated to the following result:
\begin{widetext}
\begin{equation}
\int_0^{\tau} d\tau e^{a \tau - b(\tau - \tau_0)^2} = \frac{1}{2\sqrt{b}} e^{\frac{a^2}{4b} + a\tau_0} \sqrt{\pi} \Bigg(\mathrm{Erf}\bigg(\frac{a + 2b\tau_0}{2\sqrt{b}}\bigg) + \mathrm{Erf}\bigg(\frac{-a + 2b(\tau - \tau_0)}{2\sqrt{b}}\bigg)\Bigg).
\end{equation}
\end{widetext}
Finally, we solve for the transfer fidelity. This can be accomplished either by maximizing $\beta^2(\tau)$ by taking the derivative with respect to the time $\tau$, or by solving the expression $r_{in}(\tau_{max}) = \kappa_i \beta^2(\tau_{max})$ for $\tau_{max}$, i.e. the time at which the resonator population is maximized. The latter method yields the following expression:
\begin{widetext}
\begin{align}
\begin{split}
\kappa_{max} \frac{2\sqrt{b}}{\sqrt{2\pi}} e^{-2b(\tau_{max} - \tau_0)^2} &= \kappa_i \frac{2\sqrt{b}}{\sqrt{2\pi}} e^{-(2a-1)\tau_{max}} \bigg(e^{(2a-1)\tau_c - 2b(\tau_c - \tau_0)^2} + \int_{\tau_c}^{\tau_{max}} d\tau e^{(2a-1)\tau_{max} - 2b(\tau_{max} - \tau_0)^2}\bigg), \\
e^{(2a-1)\tau_{max} - 2b(\tau_{max} - \tau_0)^2} &= (2a-1) \bigg(e^{(2a-1)\tau_c - 2b(\tau_c - \tau_0)^2} + \int_{\tau_c}^{\tau_{max}} d\tau e^{(2a-1)\tau_{max} - 2b(\tau_{max} - \tau_0)^2}\bigg),
\end{split}
\end{align}
\end{widetext}
where $\kappa_i/\kappa_{max}$ was replaced by $2a-1$. Due to the error function that results from solving the integral on the right hand side, this expression is not analytically solvable for $\tau_{max}$. As such, we again use MATLAB-based simulation, scanning over all values of $a$ and $b$, corresponding to the intrinsic loss rate and input pulse standard deviation, respectively. Recall that $\tau_c$ is also a function of these two parameters (see Eq.~\eqref{eq: t_c Gaussian}), while $\tau_0$ is a function of the standard deviation specifically, through the relationship $\tau_0 = n\kappa_{max}\sigma$. 
\par
\begin{figure}[!tb]
	\centering
	\includegraphics[width=\linewidth]{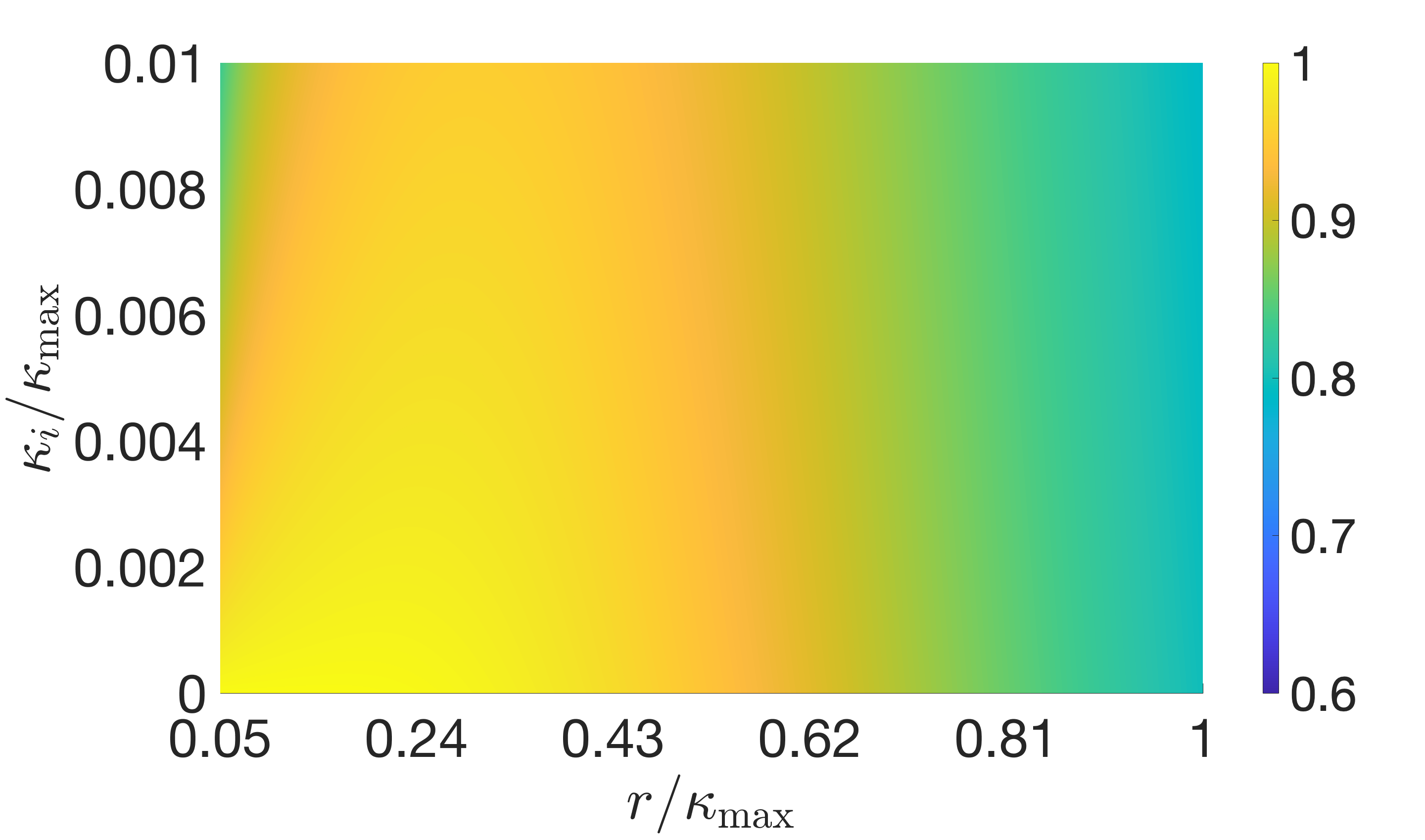}
	\caption{Fidelity $F$ as a function of intrinsic loss rate $\kappa_i$ and peak input rate $r$, both normalized with respect to the maximal achievable output coupling rate $\kappa_\mathrm{max}$. Note that the pulse standard deviation $\sigma$ relates to the peak input rate $r$ as $\sigma = 1/(r\sqrt{2\pi})$.}
	\label{fig:fidelitygaussian}
\end{figure}

 
Figure~\ref{fig:fidelitygaussian} depicts the fidelity as a function of the intrinsic loss rate $\kappa_i$ and the peak input rate $r$ (which is inversely related to the pulse width, as previously discussed) in the range $\kappa_i \le \kappa_{max}/100$ and $\kappa_{max}/20 \le r \le \kappa_{max}$, where the pulse standard deviation relates to the peak input rate as $\sigma = 1/(r\sqrt{2\pi})$. 
For the reasons discussed for the exponential input, the fidelity correlates negatively with intrinsic loss for a given pulse width, while for a given intrinsic loss rate, the fidelity is minimized for high and low pulse width values, while being maximized at a medium pulse width whose values correlates negatively with the given intrinsic loss rate.
\par
\begin{figure*}[!tb]
	\centering
	\begin{subfigure}{\columnwidth}
		\centering
		\includegraphics[width=\linewidth]{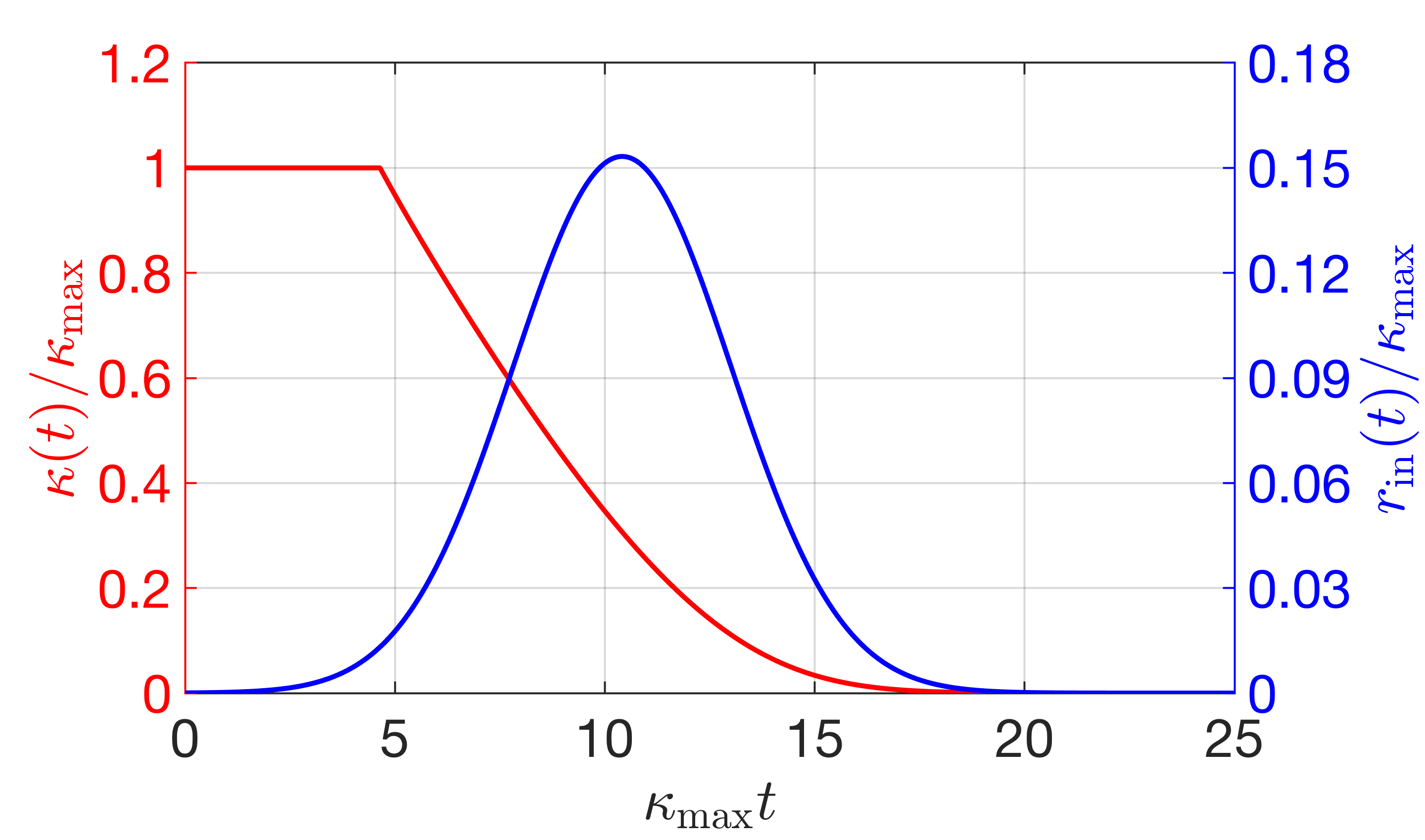}
		\caption{Input rate (blue) and optimal output coupling rate (red) vs. time.}
	\end{subfigure}
	\begin{subfigure}{\columnwidth}
		\centering
		\includegraphics[width=\linewidth]{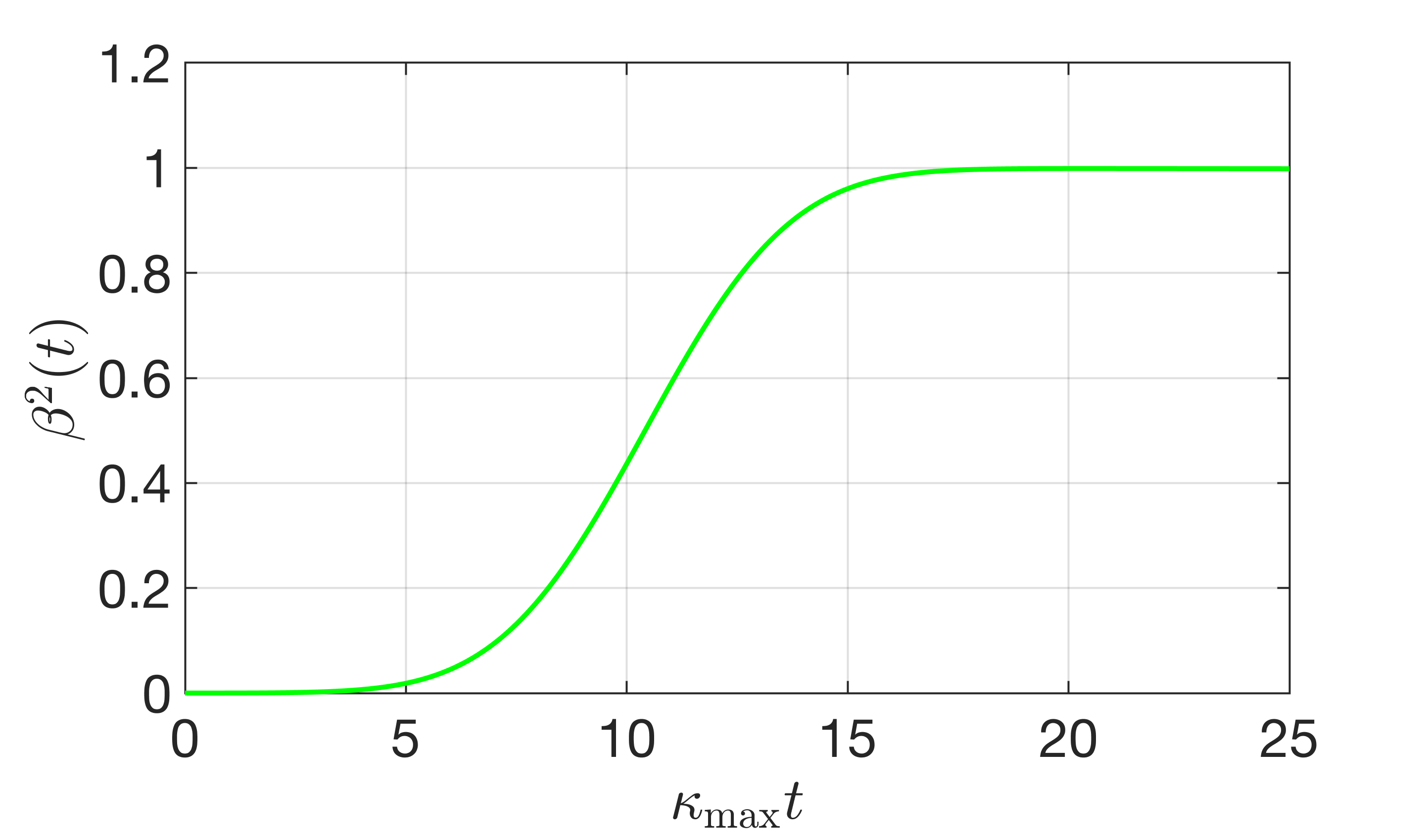}
		\caption{Resonator population vs. time.}
	\end{subfigure}
\caption{(Gaussian input case) Optimal output coupling profile $\kappa(t)$ (a), and evolution of the resonator population $\beta^2(t)$ (b), given an intrinsic loss rate $\kappa_i = \kappa_\mathrm{max}/10000$ and peak input rate $r = 0.1533 \kappa_\mathrm{max}$.}
\label{fig:gaussianinput}
\end{figure*}

 
Figure~\ref{fig:gaussianinput} depicts the input profile, output coupling profile, and resonator population time-evolution for the selected parameters $\kappa_i = \kappa_{max}/10000$ and $r = 0.1533 \kappa_{max}$. 
As in the case of the exponential input, the optimal output coupling rate steadily decreases in the second stage. From the resonator population function, we find a fidelity of 99.87\%, which is attained at time $t_{max} = 20.4/\kappa_{max}$.

\section{Conclusion}
We have derived the optimal temporal profile for the input coupling rate of a lossy resonator such that a propagating input signal carrying a qubit is maximally absorbed in a quantum memory. Furthermore, we have laid out the required time range for the short initial stage during which the net output from the quantum memory system must be temporarily non-zero, losing a small amount of quantum information over that short period of time, in order to generate a sufficient seed population in the resonator. We have also shown that the initial stage does not significantly degrade the perfect fidelity of the quantum transfer to an ideal resonator without intrinsic loss, provided that the input rate is low compared to the maximum output coupling rate. Lastly, we have calculated the fidelity of the transfer from the propagating field to the standing mode of a nonideal quantum memory resonator for two sample input field temporal profiles, namely, an exponential input and a Gaussian input, with the results showing that the fidelity varies inversely with the intrinsic loss rate, while the variation with the peak input rate is more complex, with the value of the optimal rate varying with the intrinsic loss rate. 
\par 
For a resonator with practical intrinsic loss rate and a Gaussian input rate profile, we find that the transfer fidelity can reach values as high as 99.87\%. It is worth noting that the general solution presented here is applicable to any type of quantum memory, including both the two-level fermion-field systems and the harmonic-oscillator boson-field systems, with a tunable output coupling rate, as well as any field profile for the qubit input, provided that the input field frequency is resonant with the resonator mode. In particular, we are currently in the process of building a phononic resonator consisting of a material whose output coupling rate is tuned by dynamically adjusting the strain. Coupled to an acoustic waveguide, this material will be able to absorb an incoming sound wave and store the information in its oscillatory mode. This property serves a vital role in designing a quantum memory, and along with quantum state transfer between memory blocks \cite{QSTBlocks}, it marks another milestone in building a quantum network.

In Appendix~\ref{sec: Semiclassical Solution}, we also derived the semiclassical approximation to compare with the full-quantum result. The comparison shows some similarity when the input field is real-valued, but the classical result differs from the full-quantum result for general complex input field. We attribute this discrepancy to handling the expected values semiclassically (i.e., ignoring the quantum correlations while taking the expected values). Nevertheless, the limited similarity provides an intuitive alternative understanding of the solution that is obtained through the Lagrangian formalism for the functional optimization.  
\par 
Sandia National Laboratories is a multimission laboratory managed and operated by National Technology \& Engineering Solutions of Sandia, LLC, a wholly owned subsidiary of Honeywell International Inc., for the U.S. Department of Energy’s National Nuclear Security Ad- ministration under contract DE-NA0003525.

\appendix

\section{Master Equation to Non-Hermitian Hamiltonian} \label{sec: Master Equation to Non-Hermitian Hamiltonian}

In this section, we show that given a Hilbert space spanning states with composite excitation numbers from 0 through $N$, a set of Lindblad operators consisting only of $m$-particle annihilation terms, and a beam-splitter-like Hamiltonian, the time-evolution of the Hilbert subspace consisting of states with excitation numbers higher than $N-m$ can be modeled by a non-Hermitian effective Hamiltonian. Recall that the density matrix evolves in the following manner given a Hermitian Hamiltonian $H$ and a set of Lindblad operators $L_c$, with $c$ representing a loss channel \cite{Lindblad}:
\begin{equation}
\dot{\rho} = -\frac{i}{\hbar} [H,\rho] + \sum_{c} \bigg(L_c \rho L_c^{\dag} - \frac{1}{2} L_c^{\dag} L_c \rho - \frac{1}{2} \rho L_c^{\dag} L_c\bigg).
\end{equation}
Next, we consider the individual matrix elements. We will label each row (column) using the index $i_n$ ($j_n$), where $n$ denotes the total excitation number across all modes in the system. We start by focusing on the effect of the term $L_c \rho L_c^{\dag}$ on a generic matrix element:
\begin{align}
\dot{\rho}_{i_n,j_l} &= \braket{i_n|L_c \rho L_c^{\dag}|j_l} + ... \nonumber \\
&= \braket{i'_{n+m}|\rho|j'_{l+m}} + ... \nonumber \\
&= {\rho}_{i'_{n+m},j'_{l+m}} + ...
\end{align}
Since the states with excitation number greater than $N$ are always unpopulated, and since $L_c \rho L_c^{\dag}$ raises both the row and column states by $m$, the impact of $L_c \rho L_c^{\dag}$ on the time-evolution of an element goes to zero if the excitation number for either the row state or the column state is greater than $N-m$. As a result, the matrix element for row $i_n$ and column $j_l$ given $n > N-m$ or $l > N-m$ evolves as follows:
\begin{equation} \label{eq: master equation higher elements}
\dot{\rho}_{i_n,j_l} = -\frac{i}{\hbar} [H,\rho]_{i_n,j_l} - \frac{1}{2} \sum_{c} \Big(L_c^{\dag} L_c \rho + \rho L_c^{\dag} L_c\Big)_{i_n,j_l}.
\end{equation}
Note that both the beam-splitter-like Hamiltonian $H$ and the operator $L_c^{\dag}L_c$ conserve excitation number. As such, for any excitation number greater than $N-m$, the Hilbert subspace of states with that excitation number evolves independently of other subspaces. Physically, this is due to the fact that the beam-splitter-like Hamiltonian conserves the system energy by preserving the excitation number over the entire composite system, while the Lindbladian processes degrade the energy by causing the loss of $m$ excitations to decoherence, but there exists no process that can introduce new excitations to the system.
\par
Having established that each excitation number greater than $N-m$ forms an independent subspace, we can show that its time-evolution can be modeled by an effective Hamiltonian consisting of a sum of the beam-splitter-like Hermitian $H$ and an \textit{anti}-Hermitian operator $H'$:
\begin{align}
H_\mathrm{eff} &= H + H', \nonumber \\
H_\mathrm{eff}^{\dag} &= H - H'.
\end{align}
We now consider how a generic density matrix $\rho$ evolves under this effective Hamiltonian. Recall that a density matrix is most generally defined as a weighted sum of pure states $\psi_s$, each with probability $P_s$ (the spectral decomposition theorem):
\begin{equation}
\rho(t) = \sum_{s} P_s(t) \ket{\psi_s(t)} \bra{\psi_s(t)}.
\end{equation}
Next, we take the time-derivative of $\rho$ and substitute the effective Hamiltonian via the \Schrodinger equation:
\begin{align}
\dot{\rho} &= \sum_{s} \frac{d}{dt} \Big(P_s \ket{\psi_s}\Big) \bra{\psi_s} + \sum_{s} P_s \ket{\psi_s} \bra{\dot{\psi}_s} \nonumber \\
&= -\frac{i}{\hbar} H_\mathrm{eff} \Big(\sum_{s} P_s \ket{\psi_s} \bra{\psi_s}\Big) \nonumber \\
&\quad + \Big(\sum_{s} P_s \ket{\psi_s} \bra{\psi_s}\Big) \frac{i}{\hbar} H_\mathrm{eff}^{\dag} \nonumber \\
&= -\frac{i}{\hbar} (H + H') \rho + \rho (H - H') \frac{i}{\hbar} \nonumber \\
&= -\frac{i}{\hbar} [H,\rho] - \frac{i}{\hbar} (H' \rho + \rho H').
\end{align}
Comparing this result to Eq.~\eqref{eq: master equation higher elements}, we find that the anti-Hermitian term $H'$ takes the following form in terms of the Lindblad operators:
\begin{equation}
H' = -\frac{i\hbar}{2} \sum_{c} L_c^{\dag} L_c.
\end{equation}
The fact that the master equation is equivalent to an effective non-Hermitian Hamiltonian for the states with excitation numbers greater than $N-m$ shows that if the system starts in a pure state, the Hilbert subspace of states with excitation number greater than $N-m$ remains in the pure state throughout the time-evolution, with the coefficients decaying in magnitude due to the losses from this subspace through the Lindbladian channels.
\par 
Finally, we briefly discuss the states with excitation numbers $N-m$ or lower, specifically why the corresponding matrix elements do not evolve identically under the master equation as under the effective Hamiltonian. This discrepancy is explained by the fact that these states receive the Lindblad-driven losses from the higher states, a process that generates a mixed state consisting of the original pure state and the new pure states formed from the collapse products. This mixed state phenomenon is captured by the master equation (due to the $L_c \rho L_c^{\dag}$ term) but not the effective Hamiltonian (which only models the evolution of the original pure state), and it explains why the master equation conserves the trace of the density matrix.

\section{Semiclassical Solution} \label{sec: Semiclassical Solution}

In this section, we verify the results of the full-quantum calculation using a semiclassical method, focusing on the particular case of zero intrinsic loss. We start by solving for the time evolution of the resonator annihilation operator $\dot{a}$ in the Heisenberg picture using the Heisenberg-Langevin equation of motion, which takes the following form in the rotating frame of the resonator mode:
\begin{equation}
\dot{a}(t) = -\frac{\kappa(t)}{2} a(t) - \sqrt{\kappa(t)} a_{in}(t).
\end{equation}
Subtracting $\dot{a}(t)$ from both sides, we obtain a quadratic equation in $\sqrt{\kappa(t)}$:
\begin{equation}
\sqrt{\kappa(t)} = \frac{a_{in}(t) \pm \sqrt{a_{in}^2(t) - 2a(t)\dot{a}(t)}}{a(t)}.
\end{equation}
As previously mentioned, we wish to minimize the total output field quanta over the timespan of the input-to-resonator transfer process. Since $a^{\dag}_{out}(t) a_{out}$ represents the output rate, the total output number can be found from the following integral:
\begin{equation}
N_{out} = \int_{t_i}^{t_f} dt \expect{a^{\dag}_{out}(t) a_{out}}.
\end{equation}
Substituting the input-output relationship from Eq.~\eqref{eq: input-output}, we expand the integral as follows:
\begin{align}
N_{out} &= \int_{t_i}^{t_f} dt \bigg(\expect{a^{\dag}_{in} a_{in}} + \expect{f(a,\dot{a}) \Big(a_{in}a^{\dag} + a^{\dag}_{in}a\Big)} \nonumber \\
&\quad + \expect{f^2(a,\dot{a})a^{\dag}a}\bigg),
\end{align}
where $f(a,\dot{a})$ is defined as $\sqrt{\kappa}$.
\par
Next, we use a semiclassical approximation to simplify the operators $a_{in}$, $a_{out}$ and $a$ to the respective scalar parameters $A_{in}$, $A_{out}$ and $A$, setting the expectation values of products as equivalent to the product of expectation values, such that:
\begin{align} \label{eq: output number}
N_{out} &= \int_{t_i}^{t_f} dt A^*_{out} A_{out} \nonumber \\
&= \int_{t_i}^{t_f} dt \Big(A^*_{in} A_{in} + f(A,\dot{A}) (A_{in}A^* + A^*_{in}A) \nonumber \\
&\quad + f^2(A,\dot{A}) A^*A\Big),
\end{align}
where $f(A,\dot{A})$ also becomes a scalar function, as desired since it represents the square-root of the scalar coupling rate $\kappa$:
\begin{equation} \label{eq: f(A,dot A)}
f(A,\dot{A}) = \frac{-A_{in} \pm \sqrt{A_{in}^2 - 2A\dot{A}}}{A}.
\end{equation}
In order to determine the trajectory that minimizes the integral in Eq.~\eqref{eq: output number}, it is convenient to treat the integrand as a Lagrangian and solve for $\kappa$ such that the Euler-Lagrange conditions are satisfied:
\begin{equation} \label{eq: Lagrangian}
L = A^*_{in} A_{in} + f(A,\dot{A}) (A_{in}A^* + A^*_{in}A) + f^2(A,\dot{A}) A^*A.
\end{equation}
For a given input field profile $A_{in}(t)$, it is apparent that the pairs of conjugate variables for the Lagrangian are $(A,\partial L/\partial \dot{A})$ and $(A^*,\partial L/\partial \dot{A^*})$. We start by applying Euler-Lagrange to the latter pair, since the Lagrangian is conveniently linear in $A^*$ and invariant in $\dot{A^*}$:
\begin{align}
0 &= \frac{d}{dt} \frac{\partial L}{\partial \dot{A^*}} = \frac{\partial L}{\partial A^*} \nonumber \\
&= \pm\frac{\sqrt{A_{in}^2 - 2A\dot{A}} \Big(-A_{in} \pm \sqrt{A_{in}^2 - 2A\dot{A}}\Big)}{A}.
\end{align}
This equation is solved either by setting $A_{in}^2 - 2A\dot{A} = 0$ or $\dot{A} = 0$. Physically, the latter solution actually maximizes the output field, since it corresponds to the scenario where $\kappa = 0$ and the input field is fully reflected from the resonator, with the resonator occupation number being constant over time. Instead, we focus on the former solution, which we express as the following differential equation:
\begin{equation} \label{eq: dot A}
\dot{A} = \frac{A_{in}^2}{2A}.
\end{equation}
This causes $f$ to reduce to a function of just $A$ instead of $\dot{A}$:
\begin{equation} \label{eq: f(A)}
f(A) = -\frac{A_{in}}{A}.
\end{equation}
Substituting the new expression for $f$ into Eq.~\eqref{eq: Lagrangian}, we find that the Lagrangian goes to zero:
\begin{align}
L &= A^*_{in} A_{in} + \bigg(\frac{-A_{in}}{A}\bigg) (A_{in}A^* + A^*_{in}A) \nonumber \\
&\quad + \bigg(\frac{-A_{in}}{A}\bigg)^2 A^* A \nonumber \\
&= A^*_{in} A_{in} - \frac{A_{in}^2 A^*}{A} - A^*_{in} A_{in} + \frac{A_{in}^2 A^*}{A} \nonumber \\
&= 0.
\end{align}
This result implies that, for the optimal coupling rate, the output number is actually zero, and thus the input field is fully transferred to the resonator. Note that if the Lagrangian still varied with $A$ or $\dot{A}$ even after solving the Euler-Lagrange equation for the conjugate pair $(A^*,\partial L/\partial A^*)$, then it would be necessary to solve the equation for the pair $(A,\partial L/\partial A)$ as well. However, since the Euler-Lagrange solution for $(A^*,\partial L/\partial A^*)$ yields $L = 0$, we know that the equation for $(A,\partial L/\partial A)$ is automatically satisfied. Physically, this corresponds to the fact that setting the coupling rate according to Eq.~\eqref{eq: f(A)} already leads to the global minimum value of zero for the output number, and thus further optimization is redundant.
\par
We now return to Eq.~\eqref{eq: dot A}. We multiply both sides by $2A$ and integrate in order to solve for $A(t)$ as a function of $A_{in}(t)$:
\begin{align}
A^2(t) &= \int_{0}^{A(t)} dA 2A \nonumber \\
&= A_{in}^2(t_i) + \int_{t_i}^{t} dt A_{in}^2(t).
\end{align}
From this, we find the optimal coupling rate $\kappa(t)$ by substituting into Eq.~\eqref{eq: f(A)} and squaring $f$:
\begin{equation}
\kappa(t) = \frac{A_{in}^2(t)}{A_{in}^2(t_i) + \int_{t_i}^{t} dt A_{in}^2(t)}.
\end{equation}
It is worth comparing this result to that derived from the full-quantum method. For the case of zero intrinsic loss (i.e. $\kappa_i = 0$), the denominator in the full-quantum solution is proportional to the total population of the quantum memory resonator at time $t$, i.e. $\expect{a^{\dag}(t_i)a(t_i)} + \int_{t_i}^t dt \expect{a_{in}^{\dag}(t) a_{in}(t)}$, while the numerator is proportional to the input field quanta per unit time, i.e. $\expect{a_{in}^{\dag}(t) a_{in}(t)}$. Since $A_{in}(t) = \expect{a_{in}(t)}$, it is evident that the full-quantum and semiclassical solutions align if $\expect{a_{in}(t)}^2 = \expect{a_{in}^{\dag}(t) a_{in}(t)}$.


\begin{thebibliography}{9}
	
\bibitem{Tunableresonator1}
G. D. Cole, E. S. Bjorlin, Q. Chen, C.-Y. Chan, S. Wu, C. S. Wang, N. C. MacDonald, and J. E. Bowers, MEMS-Tunable Vertical-resonator SOAs, IEEE J. Quantum Electron. \textbf{41}, 3 (2005).

\bibitem{Tunableresonator2}
H. Dong, Z. R. Gong, H. Ian, L. Zhou, and C. P. Sun, Intrinsic resonator QED and emergent quasinormal modes for a single photon, Phys. Rev. A \textbf{79}, 063847 (2009).
	
\bibitem{Tunableresonator3}
M. Tecimer, K. Holldack, and L. R. Elias, Dynamically tunable mirrors for THz free electron laser applications, Phys. Rev. ST Accel. Beams \textbf{13}, 030703 (2010).

\bibitem{Tunableresonator4}
P. Dong, A. Maho, R. Brenot, Y.-K. Chen, and A. Melikyan, Directly Reflectivity Modulated Laser, J. Lightwave Technol. \textbf{36}, 1255 (2018).

\bibitem{HarlowThesis}
J. Harlow, Microwave Electromechanics: \textit{Measuring and Manipulating the Quantum State of a Macroscopic Mechanical Oscillator,} PhD thesis, Univ. Colorado, Boulder (2013).

\bibitem{Wenner}
J. Wenner, Y. Yin, Y. Chen, R. Barends, B. Chiaro, E. Jeffrey, J. Kelly, A. Megrant, J. Y. Mutus, C. Neill, P. J. J. O’Malley, P. Roushan, D. Sank, A.
Vainsencher, T. C. White, Alexander N. Korotkov, A. N. Cleland, and J. M. Martinis, Catching Time-Reversed Microwave Coherent State Photons with 99.4\% Absorption Efficiency, Phys. Rev. Lett. \textbf{112}, 210501 (2014).

\bibitem{Nurdin}
H. I. Nurdin, M. R. James, and N. Yamamoto, Perfectly capturing traveling single photons of arbitrary temporal wavepackets with a single tunable device, arXiv:1609.05643v1.
	
\bibitem{GardinerCollett}
C. W. Gardiner and M. J. Collett, Input and output in damped quantum systems: Quantum stochastic differential equations and the master equation, Phys. Rev. A \textbf{31}, 3761 (1985).

\bibitem{BraggMirrorsMinimumTransmittance}
R. Paschotta, article on 'supermirrors' in the RP Photonics Encyclopedia, accessed on 2020-03-19.

\bibitem{SLHFormalism}
J. Combes, J. Kerckhoff, and M. Sarovar, The SLH framework for modeling quantum input-output networks, Adv. Phys.: X \textbf{2}, 784 (2017).

\bibitem{Lindblad}
G. Lindblad, On the Generators of Quantum Dynamical Semigroups, Commun. Math. Phys. \textbf{48}, 119 (1976).

\bibitem{Taylor}
J. C. Taylor, E. Chatterjee, W. Kindel, D. Soh, and M. Eichenfield, Piezo-acoustomechanical interactions for tunable quantum phononic cuircuit operation, \textit{in preparation}.

\bibitem{QSTBlocks}
D. Soh, E. Chatterjee, and M. Eichenfield, High-fidelity State Transfer Between Leaky Quantum Memories, arXiv:2005.13062v1.

\end{thebibliography}
\end{document}